\documentclass[%
 aip,
 amsmath,amssymb,
 reprint,%
]{revtex4-1}

\usepackage{graphicx}
\usepackage{dcolumn}
\usepackage{bm}

\usepackage[utf8]{inputenc}
\usepackage[T1]{fontenc}
\usepackage{mathptmx}
\usepackage{etoolbox}

\makeatletter
\def\@email#1#2{%
 \endgroup
 \patchcmd{\titleblock@produce}
  {\frontmatter@RRAPformat}
  {\frontmatter@RRAPformat{\produce@RRAP{*#1\href{mailto:#2}{#2}}}\frontmatter@RRAPformat}
  {}{}
}%

\makeatother
\begin{document}

\preprint{AIP/123-QED}
	
	\title{Ultrafast single-pulse all-optical switching in synthetic ferrimagnetic Tb/Co/Gd~multilayers}

\author{Julian Hintermayr*}%
\email{j.hintermayr@tue.nl}
\affiliation{Department of Applied Physics, Eindhoven University of Technology, P.O. Box 513, 5600 MB, Eindhoven, Netherlands}
\author{Pingzhi Li}
\affiliation{Department of Applied Physics, Eindhoven University of Technology, P.O. Box 513, 5600 MB, Eindhoven, Netherlands}
\author{Roy Rosenkamp}
\affiliation{Department of Applied Physics, Eindhoven University of Technology, P.O. Box 513, 5600 MB, Eindhoven, Netherlands}
\author{Youri L. W. van Hees}
\affiliation{Department of Applied Physics, Eindhoven University of Technology, P.O. Box 513, 5600 MB, Eindhoven, Netherlands}
\author{Junta Igarashi}
\affiliation{Université de Lorraine, Institut Jean Lamour, UMR CNRS 7198, Nancy 54011, France}
\author{Stéphane Mangin}
\affiliation{Université de Lorraine, Institut Jean Lamour, UMR CNRS 7198, Nancy 54011, France}
\author{Reinoud Lavrijsen}
\affiliation{Department of Applied Physics, Eindhoven University of Technology, P.O. Box 513, 5600 MB, Eindhoven, Netherlands}
\author{Grégory Malinowski}
\affiliation{Université de Lorraine, Institut Jean Lamour, UMR CNRS 7198, Nancy 54011, France}
\author{Bert Koopmans}
\affiliation{Department of Applied Physics, Eindhoven University of Technology, P.O. Box 513, 5600 MB, Eindhoven, Netherlands}

\begin{abstract}
    In this work, we investigate single-shot all-optical switching (AOS) in Tb/Co/Gd/Co/Tb multilayers in an attempt to establish AOS in synthetic ferrimagnets with high perpendicular magnetic anisotropy.
    In particular, we study the effect of varying Tb thicknesses to disentangle the role of the two rare-earth elements.
    Even though the role of magnetic compensation has been considered to be crucial, we find that the threshold fluence for switching is largely independent of the Tb content.
    Moreover, we identify the timescale for the magnetization to cross zero to be approximately within the first ps after laser excitation using time-resolved MOKE. We conclude that the switching is governed mostly by interactions between Co and Gd.
\end{abstract}
	
\maketitle

The energy efficiency and speed of single-shot helicity-independent all-optical switching (AOS) hold great potential for future memory and storage applications~\cite{Kimel:2019,Kim:2022aaPL}.
Following the first observation of the phenomenon of AOS in GdFeCo alloys~\cite{Ostler:2012aaPL}, 
much progress has been made towards implementing AOS in concrete memory applications.
For instance, magnetic tunnel junctions (MTJs) have been shown to be all-optically switchable by replacing the free layer with TM-rare-earth (RE) systems, using Gd-based alloys~\cite{Chen:2017}, Co/Gd multilayers~\cite{Wang:2022}, as well as [Co/Tb] multilayers~\cite{Mondal:2022}. Solutions regarding the design of photonic building blocks to address the spintronic devices in an integrated platform environment have also been demonstrated~\cite{Becker:2020aaPL,Pezeshki:2022PL,Sobolewska:2020aaPL, Pezeshki:2023}.

In all-optically switchable materials, it is deemed essential that the transition metal (TM) sublattice (Fe and Co), and the RE Gd are exchange-coupled antiparallel and exhibit a strong disparity in demagnetization timescales to enable AOS\cite{Wietstruk:2011aaPL, Kim:2022aaPL,Kimel:2019}. However, for high density data storage applications, materials based on the TM-Gd system face significant challenges, largely due to the insufficient perpendicular magnetic anisotropy (PMA) of Gd and the TM system, required to stabilize competitively small domains.

In search for materials with higher magnetic anisotropy, compounds containing the RE element Tb have since gained much attention (see Refs.~\cite{Alebrand:2012,  Hassdenteufel:2013, Mangin:2014, Hebler:2016, Lu:2018, Ciuciulkaite:2020, Krupinski:2021, Hu:2022, Liu:2022}).
Whereas this improved anisotropy is desirable from a storage density point of view, it was shown to lead to a much faster demagnetization of the $4f$ orbitals compared to Gd. Frietsch~\textit{et~al.} explained this by a transfer of angular momentum from the $4f$ moments to the magnon and phonon system which is only possible in the presence of spin--orbit coupling~\cite{Frietsch:2020}.

While AOS in TM-Gd alloys and multilayers was explained by a strong difference in demagnetization timescales between the TM and the RE---not given for TM-Tb systems---the observation of single-shot AOS in [Co/Tb] multilayers~\cite{Felix:2020, Felix:2021} came as a surprise. However, the prerequisites for AOS and the exact mechanism behind the magnetization reversal are very different. In multilayers with Tb, AOS is only observed within a small window of compositions close to magnetic compensation, whereas Co/Gd-based multilayers can be engineered to show robust switching even in samples with no magnetic compensation point~\cite{Beens:2019,Li:2023}. Another fundamental difference between the two material platforms is the timescale on which reversal occurs. In GdFeCo, it was found that the Fe magnetization crosses zero after only a few hundreds of fs~\cite{Radu:2011}, which can be explained for instance by a phenomenological description~\cite{Mentink:2012, Davies:2020, Jakobs:2022}, exchange scattering~\cite{Schellekens:2013} or nonlocal angular momentum transfer~\cite{Graves:2013}. 
Reversal in [Co/Tb] on the other hand has recently been found to take place on a much longer timescale of $\sim 100$~ps~\cite{Mishra:2023}, presumably driven by virtue of precessional switching around a transient in-plane (IP) anisotropy field, induced by ultrashort heating~\cite{Peng:2022}.
Furthermore, switching in [Co/Tb] was found to be independent of the duration of the laser stimulus in the range between 50~fs-10~ps (not present in TM-Gd systems\cite{Davies:2020}), clearly hinting towards slower switching dynamics~\cite{Peng:2022}.

Systems combining both Gd and Tb offer appealing alternatives, using Gd as the main driving force for AOS while exploiting the high magnetic anisotropy of Tb. Recently, AOS was found in systems where Gd makes up at least 20\% of the total RE material~\cite{Ceballos:2021, Zhang:2022}. A downside of alloys on the other hand are limited possibilities of (interfacial) thin film engineering compared to multilayer films.

In this work, we seek to combine the ultrafast and robust switching of Co/Gd bilayers with Tb to disentangle the roles of the two RE elements. To do this, we design multilayer stacks with Co, Gd, and Tb. We study for which RE layer thicknesses the system exhibits PMA and quantify effective anisotropy constants. Furthermore, we address the questions of the dependence of the critical fluence for AOS on magnetic compensation, which has been considered a critical parameter until now. Finally, we investigate the timescales on which this switching occurs.

\begin{figure*}[htbp]
    \centering
    \includegraphics{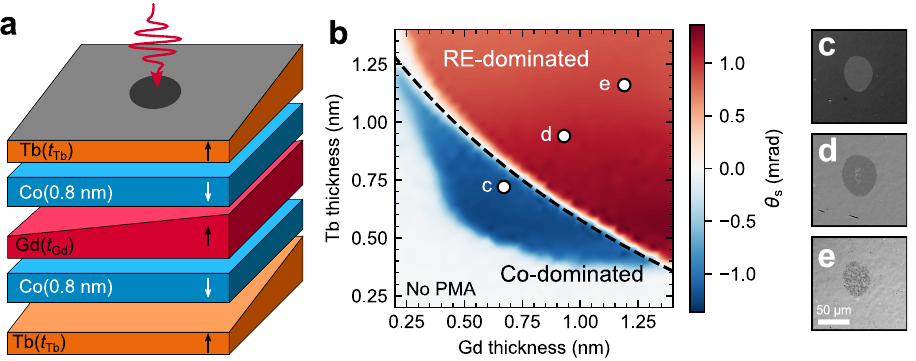}
    \caption{\textbf{a}~Sketch showing the magnetic layers within the stack with arrows indicating the local magnetization orientation. Laser excitation occurs with perpendicular incidence from free space. \textbf{b}~Phase diagram extracted from the saturation Kerr angle of the sample design shown in \textbf{a} measured with polar MOKE. The broken line indicates the compensation composition. Co-domination is indicated in blue, RE-domination in red, and IP regions in white. \textbf{c, d, e}~Kerr microscopy images of the magnetic state of the sample after excitation with a single fs laser pulse at the approximate thicknesses indicated in \textbf{b}, representing Co with magnetization up (dark) and down (light).}
    \label{fig:overview}
\end{figure*}

Samples in this study are deposited by dc magnetron sputter deposition at room temperature on Si/SiO$_2$ and doped Si:B substrates for static and time-resolved experiments, respectively.
For investigating static magnetic properties, the polar magneto-optic Kerr effect (MOKE) is used to record out-of-plane (OOP) hysteresis loops at room temperature. Further, a superconducting quantum interference device vibrating sample magnetometer (SQUID-VSM) is employed to measure IP hysteresis curves at 300~K. Single-shot all-optical switching is triggered by single linearly polarized $\sim100$~fs laser pulses at a central wavelength of 700~nm with variable energy at normal incidence. Optically reversed domains are imaged using Kerr microscopy in polar configuration. For the investigation of time-resolved dynamics, we employ time-resolved MOKE (TR-MOKE) using pump and probe pulses with a pulse duration of $<50~$fs, central wavelengths of 800~nm and 400~nm, respectively, and a repetition rate of 5~kHz.

When engineering the synthetic ferrimagnetic sample stack, it is important to consider that the Curie temperatures ($T_\mathrm{C}$) of the bulk Gd and Tb are 292~K and 222~K, respectively, and therefore below room temperature. Their magnetization will only be stabilized when interfaced with a ferromagnetic material like Co through the magnetic proximity effect, pushing $T_\mathrm{C}$ above room temperature. Taking this into account, the choice is made to perform experiments on Ta(4)/Tb($t_\mathrm{Tb}$)/Co(0.8)/Gd($t_\mathrm{Gd}$)/Co(0.8)/Tb($t_\mathrm{Tb}$)/TaN$_x$(4) multilayer samples (thicknesses in nm). A sketch of the sample design is shown in Fig.~\ref{fig:overview}~\textbf{a}.
Comments on the reproducibility of the samples are provided in the supplementary material. 

For the first sample, Gd and Tb thicknesses are spatially varied along perpendicular directions to study a wide composition range.
In order to determine the composition range where PMA is present, we measure hysteresis loops at different locations along the sample using MOKE. As the MOKE measurement is sensitive to mostly the Co magnetization, a reversal of the sign of the Kerr angle is expected when scanning from Co to RE-dominated regions. We extract the saturation Kerr angle $\theta_\mathrm{s}$ from our data and show the resulting phase diagram in Figure~\ref{fig:overview}~\textbf{b}.

\begin{figure*}[htbp]
    \centering
    \includegraphics{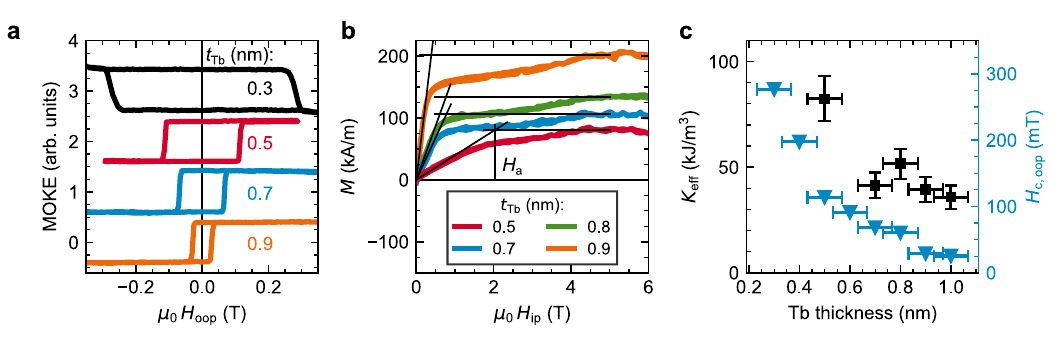}
    \caption{\textbf{a}~Perpendicular MOKE hysteresis loops (vertical offsets for clarity) and \textbf{b}~IP SQUID loops for various Tb thicknesses at a fixed Gd thickness of 1~nm measured at 300~K. A negative MOKE step in \textbf{a} indicates that the sample is Co-dominated. \textbf{c}~Effective anisotropy and coercive fields extracted from \textbf{a} and \textbf{b} as a function of Tb thickness.}
    \label{fig:SQUID}
\end{figure*}

Regions with no PMA are shown in white, Co and RE-dominated regions with PMA in blue and red, respectively. We find that PMA is absent in composition regions, where the Tb is thinner than a monolayer ($\sim 0.3~$nm). We attribute this observation to the onset of a discontinuous Co/Tb interface which is detrimental to obtaining PMA.
In regions with too thin Gd, we explain the lack of OOP anisotropy by the strong IP shape anisotropy of the joint two Co layers, overcoming interfacial anisotropy contributions.
In the region with PMA, we observe a transition from Co to RE-dominated behavior. The compensation line separating these regions is indicated by a broken line in Figure~\ref{fig:overview}~\textbf{b}.

To gain quantitative information on the effective magnetic anisotropy constant $K_\mathrm{eff}$, we record IP hysteresis loops using SQUID-VSM. For this purpose, separate samples with a fixed Gd thickness of 1~nm and varying Tb thicknesses are deposited. OOP MOKE and IP SQUID loops are presented in Fig.~\ref{fig:SQUID}~\textbf{a} and \textbf{b}, respectively. We note that the average magnetization in~\textbf{b} is calculated based on the assumption that the full volume of RE and Co layers contribute to the magnetic moment. Further elaboration on this is provided in the supplementary material. For the Co-dominated sample ($t_\mathrm{Tb}=0.3~$nm), the step of the hysteresis loop in Fig.~\ref{fig:SQUID}~\textbf{a} is inverted as discussed above. The OOP loops show well-defined square shapes with decreasing coercive fields $H_\mathrm{c}$ as $t_\mathrm{Tb}$ increases (see Fig.~\ref{fig:SQUID}~\textbf{c}). IP loops show typical hard axis reversal at low fields with close to zero remanence and a linear dependence as a function of field, until a plateau is reached. Samples with $t_\mathrm{Tb}<0.5~$nm could not be saturated along their hard axis with the available field of 7~T and corresponding loops are not shown. A peculiarity in all IP measurements is a second slow magnetization step up to $\sim4.5~$T. Such behavior has previously been ascribed to the so-called ``fanning cone'' in TM-Tb-based samples~\cite{Coey:1978, Togami:1986}.

To extract the effective uniaxial magnetic anisotropy constant from the IP data, the anisotropy field $H_\mathrm{a}$ and saturation magnetization $M_\mathrm{s}$ are determined. Due to ambiguities arising from the presence of the second step in the hysteresis loops, the following choices are made for determining $M_\mathrm{s}$ and $H_\mathrm{a}$: $M_\mathrm{s}$ is assumed to be the value of the magnetization after the second step and $H_\mathrm{a}$ is the field at which the line that is fitted to the low-field behavior intersects with $M_\mathrm{s}$. For the $t_\mathrm{Tb}=0.5~$nm sample, the extracted lines describing the low-field behavior, a horizontal line at $M_\mathrm{s}$, and a vertical line at $H_\mathrm{a}$ are indicated in Fig.~\ref{fig:SQUID}~\textbf{b}. 
The effective anisotropy constant is then given by
\begin{equation}
    K_\mathrm{eff}=\frac{1}{2}\mu_0 H_\mathrm{a} M_\mathrm{s}.
\end{equation}
Figure~\ref{fig:SQUID}~\textbf{c} shows $H_\mathrm{c}$ and $K_\mathrm{eff}$ as a function of $t_\mathrm{Tb}$. We observe a steady decrease of $K_\mathrm{eff}$ as a function of $t_\mathrm{Tb}$.
This trend could stem from a variety of effects. Firstly, the origin of the anisotropy is interfacial, meaning it is expected to scale inversely proportional to the total film thickness. Secondly, adding more Tb could result in a decrease of $T_\mathrm{C}$ which typically leads to a decrease in magnetic anisotropy. Thirdly, A higher RE content drives the material further away from compensation, leading to an increase of shape anisotropy $K_\mathrm{s}$ which decreases PMA.
Furthermore, higher thicknesses of Tb may lead to rougher Tb/Co interfaces which could also contribute to the observed decrease in PMA towards higher Tb thicknesses. Bringing the values of $K_\mathrm{eff}$ into perspective with literature data on other samples exhibiting AOS, we find anisotropy constants in a similar range as found in (Gd,Tb)Co alloys\cite{Ceballos:2021}, yet slightly lower than in Pt/Co(1~nm)/Gd(3~nm) layers\cite{Wang:2020abPL,Li:2022adPL}. We argue that it should be possible to increase $K_\mathrm{eff}$ through stack engineering, such as by adding more Tb/Co repeats.

We now turn our attention towards the characterization of AOS in our samples. Firstly, we investigate single-shot AOS at different spots in the double wedge sample, indicated in Fig.~\ref{fig:overview}~\textbf{b}. To do so, the sample is first brought to saturation by a perpendicular magnetic field. Subsequently, single fs laser pulses are directed onto the sample surface. We image the irradiated regions using Kerr microscopy and show selected results in Fig.~\ref{fig:overview}~\textbf{c}, \textbf{d}, and \textbf{e}. A more comprehensive set of Kerr images with further discussion is provided in the supplementary material. We find that in~\textbf{c} and \textbf{d}, fully switched magnetic domains are created, which is confirmed by the difference in grey level of the saturated magnetic states. It is possible to switch reversed states back to their initial state upon irradiation with subsequent laser shots.
The reversal in contrast from Fig.~\ref{fig:overview}~\textbf{c} to \textbf{d} is again caused by a transition from Co to RE dominated regions. However, for higher Gd and Tb thicknesses in Fig.~\ref{fig:overview}~\textbf{e}, only a multi-domain state is present. This is likely due to the fact that in this region, the excess of RE content brings the system too far away from magnetic compensation, preventing AOS. As discussed in Ref.~\cite{Gorchon:2016}, other reasons as to why a multidomain state may arise include the phonon temperature temporarily exceeding $T_\mathrm{C}$, or stray fields facilitating the nucleation of a multidomain state at elevated temperatures.
Indeed, a multi-domain state is already visible in Fig.~\ref{fig:overview}~\textbf{d}, although here it only occurs in the middle of the laser pulse, where the fluence is highest.

\begin{figure}[htbp]
    \centering
    \includegraphics{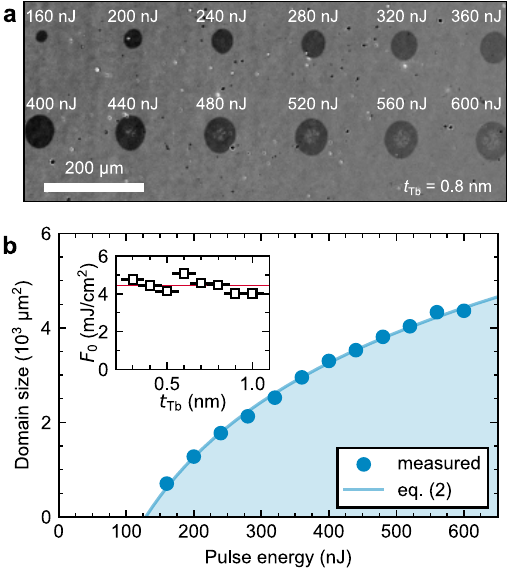}
    \caption{\textbf{a}~Kerr microscopy images of the Tb(0.8~nm)/Co/Gd(1~nm)/ Co/Tb(0.8~nm) sample after excitation by a single fs linearly polarized laser pulses with energies indicated in the Figure. \textbf{b}~Sizes of reversed domains as a function of laser pulse energy extracted from \textbf{a}. The continuous line is a best fit according to eq.~(\ref{eq:AOS}). The inset shows the critical fluence for AOS as a function of Tb thickness with the horizontal red line indicating the mean value of 4.44~mJ/cm$^2$.}
    \label{fig:domain_size}
\end{figure}

Having shown that the stack does exhibit AOS, we proceed by investigating the critical fluence for switching and the role of Tb in the process. For this experiment, we again take the sample series with fixed Gd thickness of 1~nm and different Tb thicknesses. After exposing the sample to laser pulses with different energies, the sizes of optically reversed domains are investigated. A typical image is presented in Fig.~\ref{fig:domain_size}~\textbf{a} with pulse energies annotated in the Figure.
The extracted domain size $A_\text{Domain}$ as a function of pulse energy $E$ is plotted in Fig.~\ref{fig:domain_size}~\textbf{b}. To quantify the (incident) threshold fluence $F_0$ for AOS from this measurement the equation below is used to describe the pulse energy dependence, assuming elliptical laser profiles\cite{Lalieu:2017}:
\begin{equation}\label{eq:AOS}
    A_\text{Domain}=\pi r\sigma^2\ln\left(\frac{E}{F_0\pi r\sigma^2}\right).
\end{equation}
$\sigma$ denotes the length of the ellipse, $r$ the ratio of short and long axes, and $E$ the pulse energy. The above equation is fitted to all data sets and $F_0$ is extracted, as shown in the inset of Fig.~\ref{fig:domain_size}~\textbf{b}. We find that $F_0$ is effectively independent of the Tb thickness with a mean value of 4.44~mJ/cm$^2$ at the given Gd thickness of 1~nm. This is somewhat surprising, considering the potential for AOS in pure [Co/Tb] multilayers, where AOS is observed only very close to magnetic compensation~\cite{Felix:2020, Felix:2021}. In the work on TbGdCo alloys, the threshold fluence was found to increase with increasing Tb content,~\cite{Ceballos:2021} which however also entails removal of Gd from the system.
We conclude that the switching must originate from the Co/Gd and Gd/Co interfaces which are known to facilitate ultrafast switching. It is plausible that the Tb moment is fully quenched by the laser due to its low $T_\mathrm{C}$ and high spin--orbit coupling and aligns opposite to the switched Co after remagnetization.

We note that the extended switching overview in the supplementary material suggests that for thin Gd, adding larger amounts of Tb even slightly reduces the efficiency of AOS. This notion does not contradict our findings here and could be explained by a deterioration of interface quality.

\begin{figure}[htbp]
    \centering
    \includegraphics{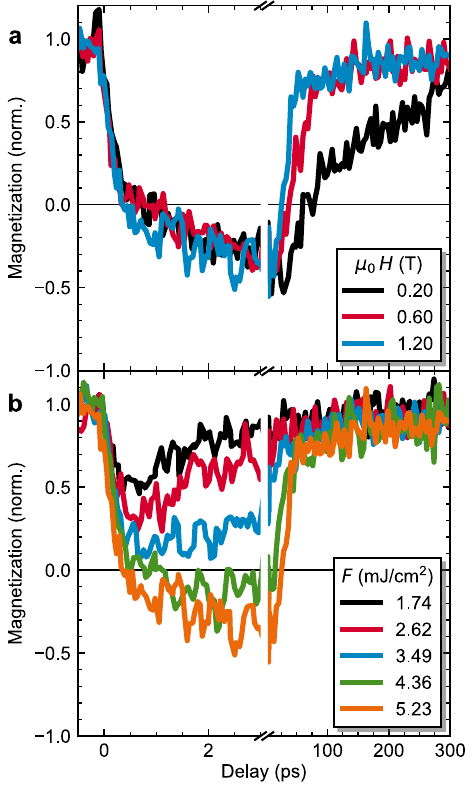}
    \caption{Time-resolved MOKE measurements of the Tb(0.3~nm)/Co/Gd(1~nm)/Co/Tb(0.3~nm) sample \textbf{a}~for different applied fields at a fluence of 5.23~mJ/cm$^{2}$ and \textbf{b}~for different laser fluences at an applied field of 1.2~T. The laser excitation occurs at a delay time of 0~ps.}
    \label{fig:TR_MOKE}
\end{figure}

Finally, we want to address the question of the time scale on which magnetization reversal takes place by measuring the time-resolved response to a fs laser pulse with TR MOKE. To ensure that the magnetization returns to its initial state after each excitation, a variable external OOP field is applied during the experiments. We investigate samples with a Gd thickness of 1~nm and a Tb thickness of 0.3~nm. Time traces for different applied field strengths are shown in Fig.~\ref{fig:TR_MOKE}~\textbf{a} at a pump fluence of 5.23~mJ/cm$^2$ which is greater than $F_0$. We find that the magnetization crosses zero approximately within the first ps after laser excitation, independently of the magnitude of the field. It is therefore not the slow precessional reversal found in [Co/Tb] multilayers, but the fast exchange-driven switching known to exist in Co/Gd systems that facilitates AOS in our samples. A full reversal to the antiparallel state is inhibited by the presence of the guiding field. 
As the magnetization slowly recovers over the following tens of ps, the magnetic state is reset by virtue of precessional switching, as found in Co/Gd~\cite{Peeters:2022}. This process is clearly accelerated by increasing the guiding field, as the measurement recorded in the lowest field of 0.2~T takes the longest time to recover its magnetization.

Furthermore, we study the dynamics at different pump fluences in Fig.~\ref{fig:TR_MOKE}~\textbf{b}. As the fluence increases, the magnetization is quenched progressively until a zero crossing occurs at a fluence of 4.36~mJ/cm$^2$. This value agrees reasonably well with $F_0=4.44~$mJ/cm$^2$, determined from static experiments. When comparing $F_0$ with well-established AOS platforms, the values are comparable to those found in pure CoGd alloys~\cite{Ceballos:2021}, but about twice as high as those in Pt/Co(1~nm)/Gd(3~nm) multilayers~\cite{Lalieu:2017}. This discrepancy could be attributed to a higher ratio of Co:Gd due to the different stack architecture.

In summary, we found PMA and robust and ultrafast AOS in Tb/Co/Gd/Co/Tb samples in a wide composition range. The zero crossing of the magnetization occurred approximately within the first ps after laser excitation, leading to the conclusion that the mechanism for reversal is identical as in Co/Gd. The fact that the addition of Tb does not result in a reduction of the critical fluence supports this notion. Our results show that hybrid structures based on Co/Gd and Co/Tb multilayers are promising candidates for magneto-photonic memory devices.

\section*{Supplementary information}
In the supplementary material, we provide further details about sample fabrication, a more comprehensive overview on AOS on the double wedge sample, and additional information on the magnetostatic characterization.

\section*{Data availability}
The data that support the findings of this study are available from the corresponding author upon reasonable request.

\section*{Conflict of Interest}
The authors have no conflicts to disclose.

\section*{Acknowledgements}
This project has received funding from the European Union’s Horizon 2020 research and innovation programme under the Marie Skłodowska-Curie grant agreement Nos.~861300 and 860060. 
This work is also part of the research program Gravitation (ZK) program "Research Center for Integrated Nanophotonics", which are financed by the Dutch Research Council (NWO). 
J.I. acknowledges support from JSPS Overseas Research Fellowships.
Furthermore, this work is supported by the ANR-20-CE09-0013 UFO, the Institute Carnot ICEEL for the project “CAPMAT” and FASTNESS, the Région Grand Est, the Metropole Grand Nancy, for the Chaire PLUS by the impact project LUE-N4S, part of the French PIA project “Lorraine Université d’Excellence” reference ANR-15-IDEX-04-LUE, the “FEDERFSE Lorraine et Massif Vosges 2014-2020” for IOMA a European Union Program, and the ANR project ANR-20-CE24-0003 SPOTZ.

\bibliography{Ref}

\begin{thebibliography}{42}%
\makeatletter
\providecommand \@ifxundefined [1]{%
 \@ifx{#1\undefined}
}%
\providecommand \@ifnum [1]{%
 \ifnum #1\expandafter \@firstoftwo
 \else \expandafter \@secondoftwo
 \fi
}%
\providecommand \@ifx [1]{%
 \ifx #1\expandafter \@firstoftwo
 \else \expandafter \@secondoftwo
 \fi
}%
\providecommand \natexlab [1]{#1}%
\providecommand \enquote  [1]{``#1''}%
\providecommand \bibnamefont  [1]{#1}%
\providecommand \bibfnamefont [1]{#1}%
\providecommand \citenamefont [1]{#1}%
\providecommand \href@noop [0]{\@secondoftwo}%
\providecommand \href [0]{\begingroup \@sanitize@url \@href}%
\providecommand \@href[1]{\@@startlink{#1}\@@href}%
\providecommand \@@href[1]{\endgroup#1\@@endlink}%
\providecommand \@sanitize@url [0]{\catcode `\\12\catcode `\$12\catcode
  `\&12\catcode `\#12\catcode `\^12\catcode `\_12\catcode `\%12\relax}%
\providecommand \@@startlink[1]{}%
\providecommand \@@endlink[0]{}%
\providecommand \url  [0]{\begingroup\@sanitize@url \@url }%
\providecommand \@url [1]{\endgroup\@href {#1}{\urlprefix }}%
\providecommand \urlprefix  [0]{URL }%
\providecommand \Eprint [0]{\href }%
\providecommand \doibase [0]{http://dx.doi.org/}%
\providecommand \selectlanguage [0]{\@gobble}%
\providecommand \bibinfo  [0]{\@secondoftwo}%
\providecommand \bibfield  [0]{\@secondoftwo}%
\providecommand \translation [1]{[#1]}%
\providecommand \BibitemOpen [0]{}%
\providecommand \bibitemStop [0]{}%
\providecommand \bibitemNoStop [0]{.\EOS\space}%
\providecommand \EOS [0]{\spacefactor3000\relax}%
\providecommand \BibitemShut  [1]{\csname bibitem#1\endcsname}%
\let\auto@bib@innerbib\@empty
\bibitem [{\citenamefont {Kimel}\ and\ \citenamefont {Li}(2019)}]{Kimel:2019}%
  \BibitemOpen
  \bibfield  {author} {\bibinfo {author} {\bibfnamefont {A.~V.}\ \bibnamefont
  {Kimel}}\ and\ \bibinfo {author} {\bibfnamefont {M.}~\bibnamefont {Li}},\
  }\bibfield  {title} {\enquote {\bibinfo {title} {Writing magnetic memory with
  ultrashort light pulses},}\ }\href {\doibase 10.1038/s41578-019-0086-3}
  {\bibfield  {journal} {\bibinfo  {journal} {Nature Reviews Materials}\
  }\textbf {\bibinfo {volume} {4}},\ \bibinfo {pages} {189} (\bibinfo {year}
  {2019})}\BibitemShut {NoStop}%
\bibitem [{\citenamefont {Kim}\ \emph {et~al.}(2022)\citenamefont {Kim},
  \citenamefont {Beach}, \citenamefont {Lee}, \citenamefont {Ono},
  \citenamefont {Rasing},\ and\ \citenamefont {Yang}}]{Kim:2022aaPL}%
  \BibitemOpen
  \bibfield  {author} {\bibinfo {author} {\bibfnamefont {S.~K.}\ \bibnamefont
  {Kim}}, \bibinfo {author} {\bibfnamefont {G.~S.~D.}\ \bibnamefont {Beach}},
  \bibinfo {author} {\bibfnamefont {K.-J.}\ \bibnamefont {Lee}}, \bibinfo
  {author} {\bibfnamefont {T.}~\bibnamefont {Ono}}, \bibinfo {author}
  {\bibfnamefont {T.}~\bibnamefont {Rasing}}, \ and\ \bibinfo {author}
  {\bibfnamefont {H.}~\bibnamefont {Yang}},\ }\bibfield  {title} {\enquote
  {\bibinfo {title} {Ferrimagnetic spintronics},}\ }\href {\doibase
  10.1038/s41563-021-01139-4} {\bibfield  {journal} {\bibinfo  {journal} {Nat.
  Mater.}\ }\textbf {\bibinfo {volume} {21}},\ \bibinfo {pages} {24} (\bibinfo
  {year} {2022})}\BibitemShut {NoStop}%
\bibitem [{\citenamefont {Ostler}\ \emph {et~al.}(2012)\citenamefont {Ostler},
  \citenamefont {Barker}, \citenamefont {Evans}, \citenamefont {Chantrell},
  \citenamefont {Atxitia}, \citenamefont {Chubykalo-Fesenko}, \citenamefont
  {El~Moussaoui}, \citenamefont {Le~Guyader}, \citenamefont {Mengotti},
  \citenamefont {Heyderman}, \citenamefont {Nolting}, \citenamefont
  {Tsukamoto}, \citenamefont {Itoh}, \citenamefont {Afanasiev}, \citenamefont
  {Ivanov}, \citenamefont {Kalashnikova}, \citenamefont {Vahaplar},
  \citenamefont {Mentink}, \citenamefont {Kirilyuk}, \citenamefont {Rasing},\
  and\ \citenamefont {Kimel}}]{Ostler:2012aaPL}%
  \BibitemOpen
  \bibfield  {author} {\bibinfo {author} {\bibfnamefont {T.~A.}\ \bibnamefont
  {Ostler}}, \bibinfo {author} {\bibfnamefont {J.}~\bibnamefont {Barker}},
  \bibinfo {author} {\bibfnamefont {R.~F.~L.}\ \bibnamefont {Evans}}, \bibinfo
  {author} {\bibfnamefont {R.~W.}\ \bibnamefont {Chantrell}}, \bibinfo {author}
  {\bibfnamefont {U.}~\bibnamefont {Atxitia}}, \bibinfo {author} {\bibfnamefont
  {O.}~\bibnamefont {Chubykalo-Fesenko}}, \bibinfo {author} {\bibfnamefont
  {S.}~\bibnamefont {El~Moussaoui}}, \bibinfo {author} {\bibfnamefont
  {L.}~\bibnamefont {Le~Guyader}}, \bibinfo {author} {\bibfnamefont
  {E.}~\bibnamefont {Mengotti}}, \bibinfo {author} {\bibfnamefont {L.~J.}\
  \bibnamefont {Heyderman}}, \bibinfo {author} {\bibfnamefont {F.}~\bibnamefont
  {Nolting}}, \bibinfo {author} {\bibfnamefont {A.}~\bibnamefont {Tsukamoto}},
  \bibinfo {author} {\bibfnamefont {A.}~\bibnamefont {Itoh}}, \bibinfo {author}
  {\bibfnamefont {D.}~\bibnamefont {Afanasiev}}, \bibinfo {author}
  {\bibfnamefont {B.~A.}\ \bibnamefont {Ivanov}}, \bibinfo {author}
  {\bibfnamefont {A.~M.}\ \bibnamefont {Kalashnikova}}, \bibinfo {author}
  {\bibfnamefont {K.}~\bibnamefont {Vahaplar}}, \bibinfo {author}
  {\bibfnamefont {J.}~\bibnamefont {Mentink}}, \bibinfo {author} {\bibfnamefont
  {A.}~\bibnamefont {Kirilyuk}}, \bibinfo {author} {\bibfnamefont
  {T.}~\bibnamefont {Rasing}}, \ and\ \bibinfo {author} {\bibfnamefont {A.~V.}\
  \bibnamefont {Kimel}},\ }\bibfield  {title} {\enquote {\bibinfo {title}
  {Ultrafast heating as a sufficient stimulus for magnetization reversal in a
  ferrimagnet},}\ }\href@noop {} {\bibfield  {journal} {\bibinfo  {journal}
  {Nature Communications}\ }\textbf {\bibinfo {volume} {3}},\ \bibinfo {pages}
  {666} (\bibinfo {year} {2012})}\BibitemShut {NoStop}%
\bibitem [{\citenamefont {Chen}\ \emph {et~al.}(2017)\citenamefont {Chen},
  \citenamefont {He}, \citenamefont {Wang},\ and\ \citenamefont
  {Li}}]{Chen:2017}%
  \BibitemOpen
  \bibfield  {author} {\bibinfo {author} {\bibfnamefont {J.-Y.}\ \bibnamefont
  {Chen}}, \bibinfo {author} {\bibfnamefont {L.}~\bibnamefont {He}}, \bibinfo
  {author} {\bibfnamefont {J.-P.}\ \bibnamefont {Wang}}, \ and\ \bibinfo
  {author} {\bibfnamefont {M.}~\bibnamefont {Li}},\ }\bibfield  {title}
  {\enquote {\bibinfo {title} {All-optical switching of magnetic tunnel
  junctions with single subpicosecond laser pulses},}\ }\href {\doibase
  10.1103/PhysRevApplied.7.021001} {\bibfield  {journal} {\bibinfo  {journal}
  {Phys. Rev. Appl.}\ }\textbf {\bibinfo {volume} {7}},\ \bibinfo {pages}
  {021001} (\bibinfo {year} {2017})}\BibitemShut {NoStop}%
\bibitem [{\citenamefont {Wang}\ \emph {et~al.}(2022)\citenamefont {Wang},
  \citenamefont {Cheng}, \citenamefont {Li}, \citenamefont {van Hees},
  \citenamefont {Liu}, \citenamefont {Cao}, \citenamefont {Lavrijsen},
  \citenamefont {Lin}, \citenamefont {Koopmans},\ and\ \citenamefont
  {Zhao}}]{Wang:2022}%
  \BibitemOpen
  \bibfield  {author} {\bibinfo {author} {\bibfnamefont {L.}~\bibnamefont
  {Wang}}, \bibinfo {author} {\bibfnamefont {H.}~\bibnamefont {Cheng}},
  \bibinfo {author} {\bibfnamefont {P.}~\bibnamefont {Li}}, \bibinfo {author}
  {\bibfnamefont {Y.~L.~W.}\ \bibnamefont {van Hees}}, \bibinfo {author}
  {\bibfnamefont {Y.}~\bibnamefont {Liu}}, \bibinfo {author} {\bibfnamefont
  {K.}~\bibnamefont {Cao}}, \bibinfo {author} {\bibfnamefont {R.}~\bibnamefont
  {Lavrijsen}}, \bibinfo {author} {\bibfnamefont {X.}~\bibnamefont {Lin}},
  \bibinfo {author} {\bibfnamefont {B.}~\bibnamefont {Koopmans}}, \ and\
  \bibinfo {author} {\bibfnamefont {W.}~\bibnamefont {Zhao}},\ }\bibfield
  {title} {\enquote {\bibinfo {title} {Picosecond optospintronic tunnel
  junctions},}\ }\href {\doibase 10.1073/pnas.2204732119} {\bibfield  {journal}
  {\bibinfo  {journal} {Proc. Natl. Acad. Sci.}\ }\textbf {\bibinfo {volume}
  {119}},\ \bibinfo {pages} {e2204732119} (\bibinfo {year} {2022})}\BibitemShut
  {NoStop}%
\bibitem [{\citenamefont {Mondal}\ \emph {et~al.}(2022)\citenamefont {Mondal},
  \citenamefont {Polley}, \citenamefont {Pattabi}, \citenamefont {Chatterjee},
  \citenamefont {Salomoni}, \citenamefont {Aviles-Felix}, \citenamefont
  {Olivier}, \citenamefont {Rubio-Roy}, \citenamefont {Diény}, \citenamefont
  {Prejbeanu}, \citenamefont {Sousa}, \citenamefont {Prejbeanu},\ and\
  \citenamefont {Bokor}}]{Mondal:2022}%
  \BibitemOpen
  \bibfield  {author} {\bibinfo {author} {\bibfnamefont {S.}~\bibnamefont
  {Mondal}}, \bibinfo {author} {\bibfnamefont {D.}~\bibnamefont {Polley}},
  \bibinfo {author} {\bibfnamefont {A.}~\bibnamefont {Pattabi}}, \bibinfo
  {author} {\bibfnamefont {J.}~\bibnamefont {Chatterjee}}, \bibinfo {author}
  {\bibfnamefont {D.}~\bibnamefont {Salomoni}}, \bibinfo {author}
  {\bibfnamefont {L.}~\bibnamefont {Aviles-Felix}}, \bibinfo {author}
  {\bibfnamefont {A.}~\bibnamefont {Olivier}}, \bibinfo {author} {\bibfnamefont
  {M.}~\bibnamefont {Rubio-Roy}}, \bibinfo {author} {\bibfnamefont
  {B.}~\bibnamefont {Diény}}, \bibinfo {author} {\bibfnamefont {L.~D.~B.}\
  \bibnamefont {Prejbeanu}}, \bibinfo {author} {\bibfnamefont {R.}~\bibnamefont
  {Sousa}}, \bibinfo {author} {\bibfnamefont {I.~L.}\ \bibnamefont
  {Prejbeanu}}, \ and\ \bibinfo {author} {\bibfnamefont {J.}~\bibnamefont
  {Bokor}},\ }\bibfield  {title} {\enquote {\bibinfo {title} {Optical switching
  in {Tb/Co}-multilayer based nanoscale magnetic tunnel junctions},}\ }\href
  {\doibase 10.48550/ARXIV.2212.10361} {\bibfield  {journal} {\bibinfo
  {journal} {arXiv}\ ,\ \bibinfo {pages} {2212.10361}} (\bibinfo {year}
  {2022})}\BibitemShut {NoStop}%
\bibitem [{\citenamefont {Becker}\ \emph {et~al.}(2020)\citenamefont {Becker},
  \citenamefont {Kr{\"u}ckel}, \citenamefont {Thourhout},\ and\ \citenamefont
  {Heck}}]{Becker:2020aaPL}%
  \BibitemOpen
  \bibfield  {author} {\bibinfo {author} {\bibfnamefont {H.}~\bibnamefont
  {Becker}}, \bibinfo {author} {\bibfnamefont {C.~J.}\ \bibnamefont
  {Kr{\"u}ckel}}, \bibinfo {author} {\bibfnamefont {D.~V.}\ \bibnamefont
  {Thourhout}}, \ and\ \bibinfo {author} {\bibfnamefont {M.~J.~R.}\
  \bibnamefont {Heck}},\ }\bibfield  {title} {\enquote {\bibinfo {title}
  {Out-of-plane focusing grating couplers for silicon photonics integration
  with optical mram technology},}\ }\href {\doibase 10.1109/JSTQE.2019.2933805}
  {\bibfield  {journal} {\bibinfo  {journal} {{IEEE} J. Sel. Top. Quantum
  Electron.}\ }\textbf {\bibinfo {volume} {26}},\ \bibinfo {pages} {1}
  (\bibinfo {year} {2020})}\BibitemShut {NoStop}%
\bibitem [{\citenamefont {Pezeshki}\ \emph {et~al.}(2022)\citenamefont
  {Pezeshki}, \citenamefont {Li}, \citenamefont {Lavrijsen}, \citenamefont
  {Tol},\ and\ \citenamefont {Koopmans}}]{Pezeshki:2022PL}%
  \BibitemOpen
  \bibfield  {author} {\bibinfo {author} {\bibfnamefont {H.}~\bibnamefont
  {Pezeshki}}, \bibinfo {author} {\bibfnamefont {P.}~\bibnamefont {Li}},
  \bibinfo {author} {\bibfnamefont {R.}~\bibnamefont {Lavrijsen}}, \bibinfo
  {author} {\bibfnamefont {J.~J. G. M. V.~D.}\ \bibnamefont {Tol}}, \ and\
  \bibinfo {author} {\bibfnamefont {B.}~\bibnamefont {Koopmans}},\ }\bibfield
  {title} {\enquote {\bibinfo {title} {Optical reading of nanoscale magnetic
  bits in an integrated photonic platform},}\ }\href {\doibase
  10.1109/jqe.2022.3224782} {\bibfield  {journal} {\bibinfo  {journal} {{IEEE}
  J. Quantum Electron.}\ ,\ \bibinfo {pages} {1}} (\bibinfo {year}
  {2022})}\BibitemShut {NoStop}%
\bibitem [{\citenamefont {Sobolewska}\ \emph {et~al.}(2020)\citenamefont
  {Sobolewska}, \citenamefont {Pelloux-Prayer}, \citenamefont {Becker},
  \citenamefont {Li}, \citenamefont {Davies}, \citenamefont {Kr{\"u}ckel},
  \citenamefont {F{\'e}lix}, \citenamefont {Olivier}, \citenamefont {Sousa},
  \citenamefont {Prejbeanu}, \citenamefont {Kiriliouk}, \citenamefont
  {Thourhout}, \citenamefont {Rasing}, \citenamefont {Moradi},\ and\
  \citenamefont {Heck}}]{Sobolewska:2020aaPL}%
  \BibitemOpen
  \bibfield  {author} {\bibinfo {author} {\bibfnamefont {E.~K.}\ \bibnamefont
  {Sobolewska}}, \bibinfo {author} {\bibfnamefont {J.}~\bibnamefont
  {Pelloux-Prayer}}, \bibinfo {author} {\bibfnamefont {H.}~\bibnamefont
  {Becker}}, \bibinfo {author} {\bibfnamefont {G.}~\bibnamefont {Li}}, \bibinfo
  {author} {\bibfnamefont {C.~S.}\ \bibnamefont {Davies}}, \bibinfo {author}
  {\bibfnamefont {C.~J.}\ \bibnamefont {Kr{\"u}ckel}}, \bibinfo {author}
  {\bibfnamefont {L.~A.}\ \bibnamefont {F{\'e}lix}}, \bibinfo {author}
  {\bibfnamefont {A.}~\bibnamefont {Olivier}}, \bibinfo {author} {\bibfnamefont
  {R.~C.}\ \bibnamefont {Sousa}}, \bibinfo {author} {\bibfnamefont {I.~L.}\
  \bibnamefont {Prejbeanu}}, \bibinfo {author} {\bibfnamefont {A.~I.}\
  \bibnamefont {Kiriliouk}}, \bibinfo {author} {\bibfnamefont {D.~V.}\
  \bibnamefont {Thourhout}}, \bibinfo {author} {\bibfnamefont {T.}~\bibnamefont
  {Rasing}}, \bibinfo {author} {\bibfnamefont {F.}~\bibnamefont {Moradi}}, \
  and\ \bibinfo {author} {\bibfnamefont {M.~J.~R.}\ \bibnamefont {Heck}},\
  }\bibfield  {title} {\enquote {\bibinfo {title} {Integration platform for
  optical switching of magnetic elements},}\ }in\ \href@noop {} {\emph
  {\bibinfo {booktitle} {Proc. SPIE}}},\ Vol.\ \bibinfo {volume} {11461}\
  (\bibinfo {year} {2020})\BibitemShut {NoStop}%
\bibitem [{\citenamefont {Pezeshki}\ \emph {et~al.}(2023)\citenamefont
  {Pezeshki}, \citenamefont {Li}, \citenamefont {Lavrijsen}, \citenamefont
  {Heck}, \citenamefont {Bente}, \citenamefont {van~der Tol},\ and\
  \citenamefont {Koopmans}}]{Pezeshki:2023}%
  \BibitemOpen
  \bibfield  {author} {\bibinfo {author} {\bibfnamefont {H.}~\bibnamefont
  {Pezeshki}}, \bibinfo {author} {\bibfnamefont {P.}~\bibnamefont {Li}},
  \bibinfo {author} {\bibfnamefont {R.}~\bibnamefont {Lavrijsen}}, \bibinfo
  {author} {\bibfnamefont {M.}~\bibnamefont {Heck}}, \bibinfo {author}
  {\bibfnamefont {E.}~\bibnamefont {Bente}}, \bibinfo {author} {\bibfnamefont
  {J.}~\bibnamefont {van~der Tol}}, \ and\ \bibinfo {author} {\bibfnamefont
  {B.}~\bibnamefont {Koopmans}},\ }\bibfield  {title} {\enquote {\bibinfo
  {title} {Integrated hybrid plasmonic-photonic device for all-optical
  switching and reading of spintronic memory},}\ }\href {\doibase
  10.1103/PhysRevApplied.19.054036} {\bibfield  {journal} {\bibinfo  {journal}
  {Phys. Rev. Appl.}\ }\textbf {\bibinfo {volume} {19}},\ \bibinfo {pages}
  {054036} (\bibinfo {year} {2023})}\BibitemShut {NoStop}%
\bibitem [{\citenamefont {Wietstruk}\ \emph {et~al.}(2011)\citenamefont
  {Wietstruk}, \citenamefont {Melnikov}, \citenamefont {Stamm}, \citenamefont
  {Kachel}, \citenamefont {Pontius}, \citenamefont {Sultan}, \citenamefont
  {Gahl}, \citenamefont {Weinelt}, \citenamefont {D{\"u}rr},\ and\
  \citenamefont {Bovensiepen}}]{Wietstruk:2011aaPL}%
  \BibitemOpen
  \bibfield  {author} {\bibinfo {author} {\bibfnamefont {M.}~\bibnamefont
  {Wietstruk}}, \bibinfo {author} {\bibfnamefont {A.}~\bibnamefont {Melnikov}},
  \bibinfo {author} {\bibfnamefont {C.}~\bibnamefont {Stamm}}, \bibinfo
  {author} {\bibfnamefont {T.}~\bibnamefont {Kachel}}, \bibinfo {author}
  {\bibfnamefont {N.}~\bibnamefont {Pontius}}, \bibinfo {author} {\bibfnamefont
  {M.}~\bibnamefont {Sultan}}, \bibinfo {author} {\bibfnamefont
  {C.}~\bibnamefont {Gahl}}, \bibinfo {author} {\bibfnamefont {M.}~\bibnamefont
  {Weinelt}}, \bibinfo {author} {\bibfnamefont {H.~A.}\ \bibnamefont
  {D{\"u}rr}}, \ and\ \bibinfo {author} {\bibfnamefont {U.}~\bibnamefont
  {Bovensiepen}},\ }\bibfield  {title} {\enquote {\bibinfo {title}
  {Hot-electron-driven enhancement of spin-lattice coupling in {Gd} and {Tb}
  $4f$ ferromagnets observed by femtosecond x-ray magnetic circular
  dichroism},}\ }\href {\doibase 10.1103/PhysRevLett.106.127401} {\bibfield
  {journal} {\bibinfo  {journal} {Phys. Rev. Lett.}\ }\textbf {\bibinfo
  {volume} {106}},\ \bibinfo {pages} {127401} (\bibinfo {year}
  {2011})}\BibitemShut {NoStop}%
\bibitem [{\citenamefont {Alebrand}\ \emph {et~al.}(2012)\citenamefont
  {Alebrand}, \citenamefont {Gottwald}, \citenamefont {Hehn}, \citenamefont
  {Steil}, \citenamefont {Cinchetti}, \citenamefont {Lacour}, \citenamefont
  {Fullerton}, \citenamefont {Aeschlimann},\ and\ \citenamefont
  {Mangin}}]{Alebrand:2012}%
  \BibitemOpen
  \bibfield  {author} {\bibinfo {author} {\bibfnamefont {S.}~\bibnamefont
  {Alebrand}}, \bibinfo {author} {\bibfnamefont {M.}~\bibnamefont {Gottwald}},
  \bibinfo {author} {\bibfnamefont {M.}~\bibnamefont {Hehn}}, \bibinfo {author}
  {\bibfnamefont {D.}~\bibnamefont {Steil}}, \bibinfo {author} {\bibfnamefont
  {M.}~\bibnamefont {Cinchetti}}, \bibinfo {author} {\bibfnamefont
  {D.}~\bibnamefont {Lacour}}, \bibinfo {author} {\bibfnamefont {E.~E.}\
  \bibnamefont {Fullerton}}, \bibinfo {author} {\bibfnamefont {M.}~\bibnamefont
  {Aeschlimann}}, \ and\ \bibinfo {author} {\bibfnamefont {S.}~\bibnamefont
  {Mangin}},\ }\bibfield  {title} {\enquote {\bibinfo {title} {{Light-induced
  magnetization reversal of high-anisotropy {TbCo} alloy films}},}\ }\href
  {\doibase 10.1063/1.4759109} {\bibfield  {journal} {\bibinfo  {journal}
  {Appl. Phys. Lett.}\ }\textbf {\bibinfo {volume} {101}},\ \bibinfo {pages}
  {162408} (\bibinfo {year} {2012})}\BibitemShut {NoStop}%
\bibitem [{\citenamefont {Hassdenteufel}\ \emph {et~al.}(2013)\citenamefont
  {Hassdenteufel}, \citenamefont {Hebler}, \citenamefont {Schubert},
  \citenamefont {Liebig}, \citenamefont {Teich}, \citenamefont {Helm},
  \citenamefont {Aeschlimann}, \citenamefont {Albrecht},\ and\ \citenamefont
  {Bratschitsch}}]{Hassdenteufel:2013}%
  \BibitemOpen
  \bibfield  {author} {\bibinfo {author} {\bibfnamefont {A.}~\bibnamefont
  {Hassdenteufel}}, \bibinfo {author} {\bibfnamefont {B.}~\bibnamefont
  {Hebler}}, \bibinfo {author} {\bibfnamefont {C.}~\bibnamefont {Schubert}},
  \bibinfo {author} {\bibfnamefont {A.}~\bibnamefont {Liebig}}, \bibinfo
  {author} {\bibfnamefont {M.}~\bibnamefont {Teich}}, \bibinfo {author}
  {\bibfnamefont {M.}~\bibnamefont {Helm}}, \bibinfo {author} {\bibfnamefont
  {M.}~\bibnamefont {Aeschlimann}}, \bibinfo {author} {\bibfnamefont
  {M.}~\bibnamefont {Albrecht}}, \ and\ \bibinfo {author} {\bibfnamefont
  {R.}~\bibnamefont {Bratschitsch}},\ }\bibfield  {title} {\enquote {\bibinfo
  {title} {Thermally assisted all-optical helicity dependent magnetic switching
  in amorphous {Fe}$_{100-x}${Tb}$_x$ alloy films},}\ }\href {\doibase
  https://doi.org/10.1002/adma.201300176} {\bibfield  {journal} {\bibinfo
  {journal} {Adv. Mater.}\ }\textbf {\bibinfo {volume} {25}},\ \bibinfo {pages}
  {3122--3128} (\bibinfo {year} {2013})},\ \Eprint
  {http://arxiv.org/abs/https://onlinelibrary.wiley.com/doi/pdf/10.1002/adma.201300176}
  {https://onlinelibrary.wiley.com/doi/pdf/10.1002/adma.201300176} \BibitemShut
  {NoStop}%
\bibitem [{\citenamefont {Mangin}\ \emph {et~al.}(2014)\citenamefont {Mangin},
  \citenamefont {Gottwald}, \citenamefont {Lambert}, \citenamefont {Steil},
  \citenamefont {Uhl{\'i}{\v{r}}}, \citenamefont {Pang}, \citenamefont {Hehn},
  \citenamefont {Alebrand}, \citenamefont {Cinchetti}, \citenamefont
  {Malinowski}, \citenamefont {Fainman}, \citenamefont {Aeschlimann},\ and\
  \citenamefont {Fullerton}}]{Mangin:2014}%
  \BibitemOpen
  \bibfield  {author} {\bibinfo {author} {\bibfnamefont {S.}~\bibnamefont
  {Mangin}}, \bibinfo {author} {\bibfnamefont {M.}~\bibnamefont {Gottwald}},
  \bibinfo {author} {\bibfnamefont {C.-H.}\ \bibnamefont {Lambert}}, \bibinfo
  {author} {\bibfnamefont {D.}~\bibnamefont {Steil}}, \bibinfo {author}
  {\bibfnamefont {V.}~\bibnamefont {Uhl{\'i}{\v{r}}}}, \bibinfo {author}
  {\bibfnamefont {L.}~\bibnamefont {Pang}}, \bibinfo {author} {\bibfnamefont
  {M.}~\bibnamefont {Hehn}}, \bibinfo {author} {\bibfnamefont {S.}~\bibnamefont
  {Alebrand}}, \bibinfo {author} {\bibfnamefont {M.}~\bibnamefont {Cinchetti}},
  \bibinfo {author} {\bibfnamefont {G.}~\bibnamefont {Malinowski}}, \bibinfo
  {author} {\bibfnamefont {Y.}~\bibnamefont {Fainman}}, \bibinfo {author}
  {\bibfnamefont {M.}~\bibnamefont {Aeschlimann}}, \ and\ \bibinfo {author}
  {\bibfnamefont {E.~E.}\ \bibnamefont {Fullerton}},\ }\bibfield  {title}
  {\enquote {\bibinfo {title} {Engineered materials for all-optical
  helicity-dependent magnetic switching},}\ }\href {\doibase 10.1038/nmat3864}
  {\bibfield  {journal} {\bibinfo  {journal} {Nat. Mater.}\ }\textbf {\bibinfo
  {volume} {13}},\ \bibinfo {pages} {286} (\bibinfo {year} {2014})}\BibitemShut
  {NoStop}%
\bibitem [{\citenamefont {Hebler}\ \emph {et~al.}(2016)\citenamefont {Hebler},
  \citenamefont {Hassdenteufel}, \citenamefont {Reinhardt}, \citenamefont
  {Karl},\ and\ \citenamefont {Albrecht}}]{Hebler:2016}%
  \BibitemOpen
  \bibfield  {author} {\bibinfo {author} {\bibfnamefont {B.}~\bibnamefont
  {Hebler}}, \bibinfo {author} {\bibfnamefont {A.}~\bibnamefont
  {Hassdenteufel}}, \bibinfo {author} {\bibfnamefont {P.}~\bibnamefont
  {Reinhardt}}, \bibinfo {author} {\bibfnamefont {H.}~\bibnamefont {Karl}}, \
  and\ \bibinfo {author} {\bibfnamefont {M.}~\bibnamefont {Albrecht}},\
  }\bibfield  {title} {\enquote {\bibinfo {title} {Ferrimagnetic {Tb–Fe}
  alloy thin films: Composition and thickness dependence of magnetic properties
  and all-optical switching},}\ }\href {\doibase 10.3389/fmats.2016.00008}
  {\bibfield  {journal} {\bibinfo  {journal} {Front. Mater.}\ }\textbf
  {\bibinfo {volume} {3}} (\bibinfo {year} {2016}),\
  10.3389/fmats.2016.00008}\BibitemShut {NoStop}%
\bibitem [{\citenamefont {Lu}\ \emph {et~al.}(2018)\citenamefont {Lu},
  \citenamefont {Zou}, \citenamefont {Hinzke}, \citenamefont {Liu},
  \citenamefont {Wang}, \citenamefont {Cheng}, \citenamefont {Wu},
  \citenamefont {Ostler}, \citenamefont {Cai}, \citenamefont {Nowak},
  \citenamefont {Chantrell}, \citenamefont {Zhai},\ and\ \citenamefont
  {Xu}}]{Lu:2018}%
  \BibitemOpen
  \bibfield  {author} {\bibinfo {author} {\bibfnamefont {X.}~\bibnamefont
  {Lu}}, \bibinfo {author} {\bibfnamefont {X.}~\bibnamefont {Zou}}, \bibinfo
  {author} {\bibfnamefont {D.}~\bibnamefont {Hinzke}}, \bibinfo {author}
  {\bibfnamefont {T.}~\bibnamefont {Liu}}, \bibinfo {author} {\bibfnamefont
  {Y.}~\bibnamefont {Wang}}, \bibinfo {author} {\bibfnamefont {T.}~\bibnamefont
  {Cheng}}, \bibinfo {author} {\bibfnamefont {J.}~\bibnamefont {Wu}}, \bibinfo
  {author} {\bibfnamefont {T.~A.}\ \bibnamefont {Ostler}}, \bibinfo {author}
  {\bibfnamefont {J.}~\bibnamefont {Cai}}, \bibinfo {author} {\bibfnamefont
  {U.}~\bibnamefont {Nowak}}, \bibinfo {author} {\bibfnamefont {R.~W.}\
  \bibnamefont {Chantrell}}, \bibinfo {author} {\bibfnamefont {Y.}~\bibnamefont
  {Zhai}}, \ and\ \bibinfo {author} {\bibfnamefont {Y.}~\bibnamefont {Xu}},\
  }\bibfield  {title} {\enquote {\bibinfo {title} {{Roles of heating and
  helicity in ultrafast all-optical magnetization switching in {TbFeCo}}},}\
  }\href {\doibase 10.1063/1.5036720} {\bibfield  {journal} {\bibinfo
  {journal} {Appl. Phys. Lett.}\ }\textbf {\bibinfo {volume} {113}},\ \bibinfo
  {pages} {032405} (\bibinfo {year} {2018})}\BibitemShut {NoStop}%
\bibitem [{\citenamefont {Ciuciulkaite}\ \emph {et~al.}(2020)\citenamefont
  {Ciuciulkaite}, \citenamefont {Mishra}, \citenamefont {Moro}, \citenamefont
  {Chioar}, \citenamefont {Rowan-Robinson}, \citenamefont {Parchenko},
  \citenamefont {Kleibert}, \citenamefont {Lindgren}, \citenamefont
  {Andersson}, \citenamefont {Davies}, \citenamefont {Kimel}, \citenamefont
  {Berritta}, \citenamefont {Oppeneer}, \citenamefont {Kirilyuk},\ and\
  \citenamefont {Kapaklis}}]{Ciuciulkaite:2020}%
  \BibitemOpen
  \bibfield  {author} {\bibinfo {author} {\bibfnamefont {A.}~\bibnamefont
  {Ciuciulkaite}}, \bibinfo {author} {\bibfnamefont {K.}~\bibnamefont
  {Mishra}}, \bibinfo {author} {\bibfnamefont {M.~V.}\ \bibnamefont {Moro}},
  \bibinfo {author} {\bibfnamefont {I.-A.}\ \bibnamefont {Chioar}}, \bibinfo
  {author} {\bibfnamefont {R.~M.}\ \bibnamefont {Rowan-Robinson}}, \bibinfo
  {author} {\bibfnamefont {S.}~\bibnamefont {Parchenko}}, \bibinfo {author}
  {\bibfnamefont {A.}~\bibnamefont {Kleibert}}, \bibinfo {author}
  {\bibfnamefont {B.}~\bibnamefont {Lindgren}}, \bibinfo {author}
  {\bibfnamefont {G.}~\bibnamefont {Andersson}}, \bibinfo {author}
  {\bibfnamefont {C.~S.}\ \bibnamefont {Davies}}, \bibinfo {author}
  {\bibfnamefont {A.}~\bibnamefont {Kimel}}, \bibinfo {author} {\bibfnamefont
  {M.}~\bibnamefont {Berritta}}, \bibinfo {author} {\bibfnamefont {P.~M.}\
  \bibnamefont {Oppeneer}}, \bibinfo {author} {\bibfnamefont {A.}~\bibnamefont
  {Kirilyuk}}, \ and\ \bibinfo {author} {\bibfnamefont {V.}~\bibnamefont
  {Kapaklis}},\ }\bibfield  {title} {\enquote {\bibinfo {title} {Magnetic and
  all-optical switching properties of amorphous {Tb}$_{x}${Co}$_{100-x}$
  alloys},}\ }\href {\doibase 10.1103/PhysRevMaterials.4.104418} {\bibfield
  {journal} {\bibinfo  {journal} {Phys. Rev. Mater.}\ }\textbf {\bibinfo
  {volume} {4}},\ \bibinfo {pages} {104418} (\bibinfo {year}
  {2020})}\BibitemShut {NoStop}%
\bibitem [{\citenamefont {Krupinski}\ \emph {et~al.}(2021)\citenamefont
  {Krupinski}, \citenamefont {Hintermayr}, \citenamefont {Sobieszczyk},\ and\
  \citenamefont {Albrecht}}]{Krupinski:2021}%
  \BibitemOpen
  \bibfield  {author} {\bibinfo {author} {\bibfnamefont {M.}~\bibnamefont
  {Krupinski}}, \bibinfo {author} {\bibfnamefont {J.}~\bibnamefont
  {Hintermayr}}, \bibinfo {author} {\bibfnamefont {P.}~\bibnamefont
  {Sobieszczyk}}, \ and\ \bibinfo {author} {\bibfnamefont {M.}~\bibnamefont
  {Albrecht}},\ }\bibfield  {title} {\enquote {\bibinfo {title} {Control of
  magnetic properties in ferrimagnetic {GdFe} and {TbFe} thin films by
  {He}$^{+}$ and {Ne}$^{+}$ irradiation},}\ }\href {\doibase
  10.1103/PhysRevMaterials.5.024405} {\bibfield  {journal} {\bibinfo  {journal}
  {Phys. Rev. Mater.}\ }\textbf {\bibinfo {volume} {5}},\ \bibinfo {pages}
  {024405} (\bibinfo {year} {2021})}\BibitemShut {NoStop}%
\bibitem [{\citenamefont {Hu}\ \emph {et~al.}(2022)\citenamefont {Hu},
  \citenamefont {Besbas}, \citenamefont {Smith}, \citenamefont {Teichert},
  \citenamefont {Atcheson}, \citenamefont {Rode}, \citenamefont {Stamenov},\
  and\ \citenamefont {Coey}}]{Hu:2022}%
  \BibitemOpen
  \bibfield  {author} {\bibinfo {author} {\bibfnamefont {Z.}~\bibnamefont
  {Hu}}, \bibinfo {author} {\bibfnamefont {J.}~\bibnamefont {Besbas}}, \bibinfo
  {author} {\bibfnamefont {R.}~\bibnamefont {Smith}}, \bibinfo {author}
  {\bibfnamefont {N.}~\bibnamefont {Teichert}}, \bibinfo {author}
  {\bibfnamefont {G.}~\bibnamefont {Atcheson}}, \bibinfo {author}
  {\bibfnamefont {K.}~\bibnamefont {Rode}}, \bibinfo {author} {\bibfnamefont
  {P.}~\bibnamefont {Stamenov}}, \ and\ \bibinfo {author} {\bibfnamefont
  {J.~M.~D.}\ \bibnamefont {Coey}},\ }\bibfield  {title} {\enquote {\bibinfo
  {title} {{Single-pulse all-optical partial switching in amorphous
  {Dy}$_x${Co}$_{1-x}$ and {Tb}$_x${Co}$_{1-x}$ with random anisotropy}},}\
  }\href {\doibase 10.1063/5.0077226} {\bibfield  {journal} {\bibinfo
  {journal} {Appl. Phys. Lett.}\ }\textbf {\bibinfo {volume} {120}},\ \bibinfo
  {pages} {112401} (\bibinfo {year} {2022})}\BibitemShut {NoStop}%
\bibitem [{\citenamefont {Liu}\ \emph {et~al.}(2023)\citenamefont {Liu},
  \citenamefont {Cheng}, \citenamefont {Vallobra}, \citenamefont {Wang},
  \citenamefont {Eimer}, \citenamefont {Zhang}, \citenamefont {Malinowski},
  \citenamefont {Hehn}, \citenamefont {Xu}, \citenamefont {Mangin},\ and\
  \citenamefont {Zhao}}]{Liu:2022}%
  \BibitemOpen
  \bibfield  {author} {\bibinfo {author} {\bibfnamefont {Y.}~\bibnamefont
  {Liu}}, \bibinfo {author} {\bibfnamefont {H.}~\bibnamefont {Cheng}}, \bibinfo
  {author} {\bibfnamefont {P.}~\bibnamefont {Vallobra}}, \bibinfo {author}
  {\bibfnamefont {H.}~\bibnamefont {Wang}}, \bibinfo {author} {\bibfnamefont
  {S.}~\bibnamefont {Eimer}}, \bibinfo {author} {\bibfnamefont
  {X.}~\bibnamefont {Zhang}}, \bibinfo {author} {\bibfnamefont
  {G.}~\bibnamefont {Malinowski}}, \bibinfo {author} {\bibfnamefont
  {M.}~\bibnamefont {Hehn}}, \bibinfo {author} {\bibfnamefont {Y.}~\bibnamefont
  {Xu}}, \bibinfo {author} {\bibfnamefont {S.}~\bibnamefont {Mangin}}, \ and\
  \bibinfo {author} {\bibfnamefont {W.}~\bibnamefont {Zhao}},\ }\bibfield
  {title} {\enquote {\bibinfo {title} {{Ultrafast single-pulse switching of
  {Tb}-dominant {CoTb} alloy}},}\ }\href {\doibase 10.1063/5.0131716}
  {\bibfield  {journal} {\bibinfo  {journal} {Appl. Phys. Lett.}\ }\textbf
  {\bibinfo {volume} {122}},\ \bibinfo {pages} {022401} (\bibinfo {year}
  {2023})}\BibitemShut {NoStop}%
\bibitem [{\citenamefont {Frietsch}\ \emph {et~al.}(2020)\citenamefont
  {Frietsch}, \citenamefont {Donges}, \citenamefont {Carley}, \citenamefont
  {Teichmann}, \citenamefont {Bowlan}, \citenamefont {Döbrich}, \citenamefont
  {Carva}, \citenamefont {Legut}, \citenamefont {Oppeneer}, \citenamefont
  {Nowak},\ and\ \citenamefont {Weinelt}}]{Frietsch:2020}%
  \BibitemOpen
  \bibfield  {author} {\bibinfo {author} {\bibfnamefont {B.}~\bibnamefont
  {Frietsch}}, \bibinfo {author} {\bibfnamefont {A.}~\bibnamefont {Donges}},
  \bibinfo {author} {\bibfnamefont {R.}~\bibnamefont {Carley}}, \bibinfo
  {author} {\bibfnamefont {M.}~\bibnamefont {Teichmann}}, \bibinfo {author}
  {\bibfnamefont {J.}~\bibnamefont {Bowlan}}, \bibinfo {author} {\bibfnamefont
  {K.}~\bibnamefont {Döbrich}}, \bibinfo {author} {\bibfnamefont
  {K.}~\bibnamefont {Carva}}, \bibinfo {author} {\bibfnamefont
  {D.}~\bibnamefont {Legut}}, \bibinfo {author} {\bibfnamefont {P.~M.}\
  \bibnamefont {Oppeneer}}, \bibinfo {author} {\bibfnamefont {U.}~\bibnamefont
  {Nowak}}, \ and\ \bibinfo {author} {\bibfnamefont {M.}~\bibnamefont
  {Weinelt}},\ }\bibfield  {title} {\enquote {\bibinfo {title} {The role of
  ultrafast magnon generation in the magnetization dynamics of rare-earth
  metals},}\ }\href {\doibase 10.1126/sciadv.abb1601} {\bibfield  {journal}
  {\bibinfo  {journal} {Sci. Adv.}\ }\textbf {\bibinfo {volume} {6}},\ \bibinfo
  {pages} {eabb1601} (\bibinfo {year} {2020})}\BibitemShut {NoStop}%
\bibitem [{\citenamefont {Avil{\'e}s-F{\'e}lix}\ \emph
  {et~al.}(2020)\citenamefont {Avil{\'e}s-F{\'e}lix}, \citenamefont {Olivier},
  \citenamefont {Li}, \citenamefont {Davies}, \citenamefont
  {{\'A}lvaro-G{\'o}mez}, \citenamefont {Rubio-Roy}, \citenamefont {Auffret},
  \citenamefont {Kirilyuk}, \citenamefont {Kimel}, \citenamefont {Rasing},
  \citenamefont {Buda-Prejbeanu}, \citenamefont {Sousa}, \citenamefont
  {Dieny},\ and\ \citenamefont {Prejbeanu}}]{Felix:2020}%
  \BibitemOpen
  \bibfield  {author} {\bibinfo {author} {\bibfnamefont {L.}~\bibnamefont
  {Avil{\'e}s-F{\'e}lix}}, \bibinfo {author} {\bibfnamefont {A.}~\bibnamefont
  {Olivier}}, \bibinfo {author} {\bibfnamefont {G.}~\bibnamefont {Li}},
  \bibinfo {author} {\bibfnamefont {C.~S.}\ \bibnamefont {Davies}}, \bibinfo
  {author} {\bibfnamefont {L.}~\bibnamefont {{\'A}lvaro-G{\'o}mez}}, \bibinfo
  {author} {\bibfnamefont {M.}~\bibnamefont {Rubio-Roy}}, \bibinfo {author}
  {\bibfnamefont {S.}~\bibnamefont {Auffret}}, \bibinfo {author} {\bibfnamefont
  {A.}~\bibnamefont {Kirilyuk}}, \bibinfo {author} {\bibfnamefont {A.~V.}\
  \bibnamefont {Kimel}}, \bibinfo {author} {\bibfnamefont {T.}~\bibnamefont
  {Rasing}}, \bibinfo {author} {\bibfnamefont {L.~D.}\ \bibnamefont
  {Buda-Prejbeanu}}, \bibinfo {author} {\bibfnamefont {R.~C.}\ \bibnamefont
  {Sousa}}, \bibinfo {author} {\bibfnamefont {B.}~\bibnamefont {Dieny}}, \ and\
  \bibinfo {author} {\bibfnamefont {I.~L.}\ \bibnamefont {Prejbeanu}},\
  }\bibfield  {title} {\enquote {\bibinfo {title} {Single-shot all-optical
  switching of magnetization in {Tb}/{Co} multilayer-based electrodes},}\
  }\href {\doibase 10.1038/s41598-020-62104-w} {\bibfield  {journal} {\bibinfo
  {journal} {Sci. Rep.}\ }\textbf {\bibinfo {volume} {10}},\ \bibinfo {pages}
  {5211} (\bibinfo {year} {2020})}\BibitemShut {NoStop}%
\bibitem [{\citenamefont {Avil{\'e}s-F{\'e}lix}\ \emph
  {et~al.}(2021)\citenamefont {Avil{\'e}s-F{\'e}lix}, \citenamefont {Farcis},
  \citenamefont {Jin}, \citenamefont {{\'A}lvaro-G{\'o}mez}, \citenamefont
  {Li}, \citenamefont {Yamada}, \citenamefont {Kirilyuk}, \citenamefont
  {Kimel}, \citenamefont {Rasing}, \citenamefont {Dieny}, \citenamefont
  {Sousa}, \citenamefont {Prejbeanu},\ and\ \citenamefont
  {Buda-Prejbeanu}}]{Felix:2021}%
  \BibitemOpen
  \bibfield  {author} {\bibinfo {author} {\bibfnamefont {L.}~\bibnamefont
  {Avil{\'e}s-F{\'e}lix}}, \bibinfo {author} {\bibfnamefont {L.}~\bibnamefont
  {Farcis}}, \bibinfo {author} {\bibfnamefont {Z.}~\bibnamefont {Jin}},
  \bibinfo {author} {\bibfnamefont {L.}~\bibnamefont {{\'A}lvaro-G{\'o}mez}},
  \bibinfo {author} {\bibfnamefont {G.}~\bibnamefont {Li}}, \bibinfo {author}
  {\bibfnamefont {K.~T.}\ \bibnamefont {Yamada}}, \bibinfo {author}
  {\bibfnamefont {A.}~\bibnamefont {Kirilyuk}}, \bibinfo {author}
  {\bibfnamefont {A.~V.}\ \bibnamefont {Kimel}}, \bibinfo {author}
  {\bibfnamefont {T.}~\bibnamefont {Rasing}}, \bibinfo {author} {\bibfnamefont
  {B.}~\bibnamefont {Dieny}}, \bibinfo {author} {\bibfnamefont {R.~C.}\
  \bibnamefont {Sousa}}, \bibinfo {author} {\bibfnamefont {I.~L.}\ \bibnamefont
  {Prejbeanu}}, \ and\ \bibinfo {author} {\bibfnamefont {L.~D.}\ \bibnamefont
  {Buda-Prejbeanu}},\ }\bibfield  {title} {\enquote {\bibinfo {title}
  {All-optical spin switching probability in [{Tb}/{Co}] multilayers},}\ }\href
  {\doibase 10.1038/s41598-021-86065-w} {\bibfield  {journal} {\bibinfo
  {journal} {Sci. Rep.}\ }\textbf {\bibinfo {volume} {11}},\ \bibinfo {pages}
  {6576} (\bibinfo {year} {2021})}\BibitemShut {NoStop}%
\bibitem [{\citenamefont {Beens}\ \emph {et~al.}(2019)\citenamefont {Beens},
  \citenamefont {Lalieu}, \citenamefont {Deenen}, \citenamefont {Duine},\ and\
  \citenamefont {Koopmans}}]{Beens:2019}%
  \BibitemOpen
  \bibfield  {author} {\bibinfo {author} {\bibfnamefont {M.}~\bibnamefont
  {Beens}}, \bibinfo {author} {\bibfnamefont {M.~L.~M.}\ \bibnamefont
  {Lalieu}}, \bibinfo {author} {\bibfnamefont {A.~J.~M.}\ \bibnamefont
  {Deenen}}, \bibinfo {author} {\bibfnamefont {R.~A.}\ \bibnamefont {Duine}}, \
  and\ \bibinfo {author} {\bibfnamefont {B.}~\bibnamefont {Koopmans}},\
  }\bibfield  {title} {\enquote {\bibinfo {title} {Comparing all-optical
  switching in synthetic-ferrimagnetic multilayers and alloys},}\ }\href
  {\doibase 10.1103/PhysRevB.100.220409} {\bibfield  {journal} {\bibinfo
  {journal} {Phys. Rev. B}\ }\textbf {\bibinfo {volume} {100}},\ \bibinfo
  {pages} {220409} (\bibinfo {year} {2019})}\BibitemShut {NoStop}%
\bibitem [{\citenamefont {Li}\ \emph {et~al.}(2023)\citenamefont {Li},
  \citenamefont {Kools}, \citenamefont {Koopmans},\ and\ \citenamefont
  {Lavrijsen}}]{Li:2023}%
  \BibitemOpen
  \bibfield  {author} {\bibinfo {author} {\bibfnamefont {P.}~\bibnamefont
  {Li}}, \bibinfo {author} {\bibfnamefont {T.~J.}\ \bibnamefont {Kools}},
  \bibinfo {author} {\bibfnamefont {B.}~\bibnamefont {Koopmans}}, \ and\
  \bibinfo {author} {\bibfnamefont {R.}~\bibnamefont {Lavrijsen}},\ }\bibfield
  {title} {\enquote {\bibinfo {title} {Ultrafast racetrack based on compensated
  {Co/Gd}-based synthetic ferrimagnet with all-optical switching},}\ }\href
  {\doibase https://doi.org/10.1002/aelm.202200613} {\bibfield  {journal}
  {\bibinfo  {journal} {Adv. Electron. Mater.}\ }\textbf {\bibinfo {volume}
  {9}},\ \bibinfo {pages} {2200613} (\bibinfo {year} {2023})}\BibitemShut
  {NoStop}%
\bibitem [{\citenamefont {Radu}\ \emph {et~al.}(2011)\citenamefont {Radu},
  \citenamefont {Vahaplar}, \citenamefont {Stamm}, \citenamefont {Kachel},
  \citenamefont {Pontius}, \citenamefont {D{\"u}rr}, \citenamefont {Ostler},
  \citenamefont {Barker}, \citenamefont {Evans}, \citenamefont {Chantrell},
  \citenamefont {Tsukamoto}, \citenamefont {Itoh}, \citenamefont {Kirilyuk},
  \citenamefont {Rasing},\ and\ \citenamefont {Kimel}}]{Radu:2011}%
  \BibitemOpen
  \bibfield  {author} {\bibinfo {author} {\bibfnamefont {I.}~\bibnamefont
  {Radu}}, \bibinfo {author} {\bibfnamefont {K.}~\bibnamefont {Vahaplar}},
  \bibinfo {author} {\bibfnamefont {C.}~\bibnamefont {Stamm}}, \bibinfo
  {author} {\bibfnamefont {T.}~\bibnamefont {Kachel}}, \bibinfo {author}
  {\bibfnamefont {N.}~\bibnamefont {Pontius}}, \bibinfo {author} {\bibfnamefont
  {H.~A.}\ \bibnamefont {D{\"u}rr}}, \bibinfo {author} {\bibfnamefont {T.~A.}\
  \bibnamefont {Ostler}}, \bibinfo {author} {\bibfnamefont {J.}~\bibnamefont
  {Barker}}, \bibinfo {author} {\bibfnamefont {R.~F.~L.}\ \bibnamefont
  {Evans}}, \bibinfo {author} {\bibfnamefont {R.~W.}\ \bibnamefont
  {Chantrell}}, \bibinfo {author} {\bibfnamefont {A.}~\bibnamefont
  {Tsukamoto}}, \bibinfo {author} {\bibfnamefont {A.}~\bibnamefont {Itoh}},
  \bibinfo {author} {\bibfnamefont {A.}~\bibnamefont {Kirilyuk}}, \bibinfo
  {author} {\bibfnamefont {T.}~\bibnamefont {Rasing}}, \ and\ \bibinfo {author}
  {\bibfnamefont {A.~V.}\ \bibnamefont {Kimel}},\ }\bibfield  {title} {\enquote
  {\bibinfo {title} {Transient ferromagnetic-like state mediating ultrafast
  reversal of antiferromagnetically coupled spins},}\ }\href {\doibase
  10.1038/nature09901} {\bibfield  {journal} {\bibinfo  {journal} {Nature}\
  }\textbf {\bibinfo {volume} {472}},\ \bibinfo {pages} {205} (\bibinfo {year}
  {2011})}\BibitemShut {NoStop}%
\bibitem [{\citenamefont {Mentink}\ \emph {et~al.}(2012)\citenamefont
  {Mentink}, \citenamefont {Hellsvik}, \citenamefont {Afanasiev}, \citenamefont
  {Ivanov}, \citenamefont {Kirilyuk}, \citenamefont {Kimel}, \citenamefont
  {Eriksson}, \citenamefont {Katsnelson},\ and\ \citenamefont
  {Rasing}}]{Mentink:2012}%
  \BibitemOpen
  \bibfield  {author} {\bibinfo {author} {\bibfnamefont {J.~H.}\ \bibnamefont
  {Mentink}}, \bibinfo {author} {\bibfnamefont {J.}~\bibnamefont {Hellsvik}},
  \bibinfo {author} {\bibfnamefont {D.~V.}\ \bibnamefont {Afanasiev}}, \bibinfo
  {author} {\bibfnamefont {B.~A.}\ \bibnamefont {Ivanov}}, \bibinfo {author}
  {\bibfnamefont {A.}~\bibnamefont {Kirilyuk}}, \bibinfo {author}
  {\bibfnamefont {A.~V.}\ \bibnamefont {Kimel}}, \bibinfo {author}
  {\bibfnamefont {O.}~\bibnamefont {Eriksson}}, \bibinfo {author}
  {\bibfnamefont {M.~I.}\ \bibnamefont {Katsnelson}}, \ and\ \bibinfo {author}
  {\bibfnamefont {T.}~\bibnamefont {Rasing}},\ }\bibfield  {title} {\enquote
  {\bibinfo {title} {Ultrafast spin dynamics in multisublattice magnets},}\
  }\href {\doibase 10.1103/PhysRevLett.108.057202} {\bibfield  {journal}
  {\bibinfo  {journal} {Phys. Rev. Lett.}\ }\textbf {\bibinfo {volume} {108}},\
  \bibinfo {pages} {057202} (\bibinfo {year} {2012})}\BibitemShut {NoStop}%
\bibitem [{\citenamefont {Davies}\ \emph {et~al.}(2020)\citenamefont {Davies},
  \citenamefont {Janssen}, \citenamefont {Mentink}, \citenamefont {Tsukamoto},
  \citenamefont {Kimel}, \citenamefont {van~der Meer}, \citenamefont
  {Stupakiewicz},\ and\ \citenamefont {Kirilyuk}}]{Davies:2020}%
  \BibitemOpen
  \bibfield  {author} {\bibinfo {author} {\bibfnamefont {C.}~\bibnamefont
  {Davies}}, \bibinfo {author} {\bibfnamefont {T.}~\bibnamefont {Janssen}},
  \bibinfo {author} {\bibfnamefont {J.}~\bibnamefont {Mentink}}, \bibinfo
  {author} {\bibfnamefont {A.}~\bibnamefont {Tsukamoto}}, \bibinfo {author}
  {\bibfnamefont {A.}~\bibnamefont {Kimel}}, \bibinfo {author} {\bibfnamefont
  {A.}~\bibnamefont {van~der Meer}}, \bibinfo {author} {\bibfnamefont
  {A.}~\bibnamefont {Stupakiewicz}}, \ and\ \bibinfo {author} {\bibfnamefont
  {A.}~\bibnamefont {Kirilyuk}},\ }\bibfield  {title} {\enquote {\bibinfo
  {title} {Pathways for single-shot all-optical switching of magnetization in
  ferrimagnets},}\ }\href {\doibase 10.1103/PhysRevApplied.13.024064}
  {\bibfield  {journal} {\bibinfo  {journal} {Phys. Rev. Appl.}\ }\textbf
  {\bibinfo {volume} {13}},\ \bibinfo {pages} {024064} (\bibinfo {year}
  {2020})}\BibitemShut {NoStop}%
\bibitem [{\citenamefont {Jakobs}\ and\ \citenamefont
  {Atxitia}(2022)}]{Jakobs:2022}%
  \BibitemOpen
  \bibfield  {author} {\bibinfo {author} {\bibfnamefont {F.}~\bibnamefont
  {Jakobs}}\ and\ \bibinfo {author} {\bibfnamefont {U.}~\bibnamefont
  {Atxitia}},\ }\bibfield  {title} {\enquote {\bibinfo {title} {Universal
  criteria for single femtosecond pulse ultrafast magnetization switching in
  ferrimagnets},}\ }\href {\doibase 10.1103/PhysRevLett.129.037203} {\bibfield
  {journal} {\bibinfo  {journal} {Phys. Rev. Lett.}\ }\textbf {\bibinfo
  {volume} {129}},\ \bibinfo {pages} {037203} (\bibinfo {year}
  {2022})}\BibitemShut {NoStop}%
\bibitem [{\citenamefont {Schellekens}\ and\ \citenamefont
  {Koopmans}(2013)}]{Schellekens:2013}%
  \BibitemOpen
  \bibfield  {author} {\bibinfo {author} {\bibfnamefont {A.~J.}\ \bibnamefont
  {Schellekens}}\ and\ \bibinfo {author} {\bibfnamefont {B.}~\bibnamefont
  {Koopmans}},\ }\bibfield  {title} {\enquote {\bibinfo {title} {Microscopic
  model for ultrafast magnetization dynamics of multisublattice magnets},}\
  }\href {\doibase 10.1103/PhysRevB.87.020407} {\bibfield  {journal} {\bibinfo
  {journal} {Phys. Rev. B}\ }\textbf {\bibinfo {volume} {87}},\ \bibinfo
  {pages} {020407} (\bibinfo {year} {2013})}\BibitemShut {NoStop}%
\bibitem [{\citenamefont {Graves}\ \emph {et~al.}(2013)\citenamefont {Graves},
  \citenamefont {Reid}, \citenamefont {Wang}, \citenamefont {Wu}, \citenamefont
  {de~Jong}, \citenamefont {Vahaplar}, \citenamefont {Radu}, \citenamefont
  {Bernstein}, \citenamefont {Messerschmidt}, \citenamefont {M{\"u}ller},
  \citenamefont {Coffee}, \citenamefont {Bionta}, \citenamefont {Epp},
  \citenamefont {Hartmann}, \citenamefont {Kimmel}, \citenamefont {Hauser},
  \citenamefont {Hartmann}, \citenamefont {Holl}, \citenamefont {Gorke},
  \citenamefont {Mentink}, \citenamefont {Tsukamoto}, \citenamefont {Fognini},
  \citenamefont {Turner}, \citenamefont {Schlotter}, \citenamefont {Rolles},
  \citenamefont {Soltau}, \citenamefont {Str{\"u}der}, \citenamefont
  {Acremann}, \citenamefont {Kimel}, \citenamefont {Kirilyuk}, \citenamefont
  {Rasing}, \citenamefont {St{\"o}hr}, \citenamefont {Scherz},\ and\
  \citenamefont {D{\"u}rr}}]{Graves:2013}%
  \BibitemOpen
  \bibfield  {author} {\bibinfo {author} {\bibfnamefont {C.~E.}\ \bibnamefont
  {Graves}}, \bibinfo {author} {\bibfnamefont {A.~H.}\ \bibnamefont {Reid}},
  \bibinfo {author} {\bibfnamefont {T.}~\bibnamefont {Wang}}, \bibinfo {author}
  {\bibfnamefont {B.}~\bibnamefont {Wu}}, \bibinfo {author} {\bibfnamefont
  {S.}~\bibnamefont {de~Jong}}, \bibinfo {author} {\bibfnamefont
  {K.}~\bibnamefont {Vahaplar}}, \bibinfo {author} {\bibfnamefont
  {I.}~\bibnamefont {Radu}}, \bibinfo {author} {\bibfnamefont {D.~P.}\
  \bibnamefont {Bernstein}}, \bibinfo {author} {\bibfnamefont {M.}~\bibnamefont
  {Messerschmidt}}, \bibinfo {author} {\bibfnamefont {L.}~\bibnamefont
  {M{\"u}ller}}, \bibinfo {author} {\bibfnamefont {R.}~\bibnamefont {Coffee}},
  \bibinfo {author} {\bibfnamefont {M.}~\bibnamefont {Bionta}}, \bibinfo
  {author} {\bibfnamefont {S.~W.}\ \bibnamefont {Epp}}, \bibinfo {author}
  {\bibfnamefont {R.}~\bibnamefont {Hartmann}}, \bibinfo {author}
  {\bibfnamefont {N.}~\bibnamefont {Kimmel}}, \bibinfo {author} {\bibfnamefont
  {G.}~\bibnamefont {Hauser}}, \bibinfo {author} {\bibfnamefont
  {A.}~\bibnamefont {Hartmann}}, \bibinfo {author} {\bibfnamefont
  {P.}~\bibnamefont {Holl}}, \bibinfo {author} {\bibfnamefont {H.}~\bibnamefont
  {Gorke}}, \bibinfo {author} {\bibfnamefont {J.~H.}\ \bibnamefont {Mentink}},
  \bibinfo {author} {\bibfnamefont {A.}~\bibnamefont {Tsukamoto}}, \bibinfo
  {author} {\bibfnamefont {A.}~\bibnamefont {Fognini}}, \bibinfo {author}
  {\bibfnamefont {J.~J.}\ \bibnamefont {Turner}}, \bibinfo {author}
  {\bibfnamefont {W.~F.}\ \bibnamefont {Schlotter}}, \bibinfo {author}
  {\bibfnamefont {D.}~\bibnamefont {Rolles}}, \bibinfo {author} {\bibfnamefont
  {H.}~\bibnamefont {Soltau}}, \bibinfo {author} {\bibfnamefont
  {L.}~\bibnamefont {Str{\"u}der}}, \bibinfo {author} {\bibfnamefont
  {Y.}~\bibnamefont {Acremann}}, \bibinfo {author} {\bibfnamefont {A.~V.}\
  \bibnamefont {Kimel}}, \bibinfo {author} {\bibfnamefont {A.}~\bibnamefont
  {Kirilyuk}}, \bibinfo {author} {\bibfnamefont {T.}~\bibnamefont {Rasing}},
  \bibinfo {author} {\bibfnamefont {J.}~\bibnamefont {St{\"o}hr}}, \bibinfo
  {author} {\bibfnamefont {A.~O.}\ \bibnamefont {Scherz}}, \ and\ \bibinfo
  {author} {\bibfnamefont {H.~A.}\ \bibnamefont {D{\"u}rr}},\ }\bibfield
  {title} {\enquote {\bibinfo {title} {Nanoscale spin reversal by non-local
  angular momentum transfer following ultrafast laser excitation in
  ferrimagnetic {GdFeCo}},}\ }\href {\doibase 10.1038/nmat3597} {\bibfield
  {journal} {\bibinfo  {journal} {Nat. Mater.}\ }\textbf {\bibinfo {volume}
  {12}},\ \bibinfo {pages} {293} (\bibinfo {year} {2013})}\BibitemShut
  {NoStop}%
\bibitem [{\citenamefont {Mishra}\ \emph {et~al.}(2023)\citenamefont {Mishra},
  \citenamefont {Blank}, \citenamefont {Davies}, \citenamefont
  {Avil\'es-F\'elix}, \citenamefont {Salomoni}, \citenamefont {Buda-Prejbeanu},
  \citenamefont {Sousa}, \citenamefont {Prejbeanu}, \citenamefont {Koopmans},
  \citenamefont {Rasing}, \citenamefont {Kimel},\ and\ \citenamefont
  {Kirilyuk}}]{Mishra:2023}%
  \BibitemOpen
  \bibfield  {author} {\bibinfo {author} {\bibfnamefont {K.}~\bibnamefont
  {Mishra}}, \bibinfo {author} {\bibfnamefont {T.~G.~H.}\ \bibnamefont
  {Blank}}, \bibinfo {author} {\bibfnamefont {C.~S.}\ \bibnamefont {Davies}},
  \bibinfo {author} {\bibfnamefont {L.}~\bibnamefont {Avil\'es-F\'elix}},
  \bibinfo {author} {\bibfnamefont {D.}~\bibnamefont {Salomoni}}, \bibinfo
  {author} {\bibfnamefont {L.~D.}\ \bibnamefont {Buda-Prejbeanu}}, \bibinfo
  {author} {\bibfnamefont {R.~C.}\ \bibnamefont {Sousa}}, \bibinfo {author}
  {\bibfnamefont {I.~L.}\ \bibnamefont {Prejbeanu}}, \bibinfo {author}
  {\bibfnamefont {B.}~\bibnamefont {Koopmans}}, \bibinfo {author}
  {\bibfnamefont {T.}~\bibnamefont {Rasing}}, \bibinfo {author} {\bibfnamefont
  {A.~V.}\ \bibnamefont {Kimel}}, \ and\ \bibinfo {author} {\bibfnamefont
  {A.}~\bibnamefont {Kirilyuk}},\ }\bibfield  {title} {\enquote {\bibinfo
  {title} {Dynamics of all-optical single-shot switching of magnetization in
  {Tb/Co} multilayers},}\ }\href {\doibase 10.1103/PhysRevResearch.5.023163}
  {\bibfield  {journal} {\bibinfo  {journal} {Phys. Rev. Res.}\ }\textbf
  {\bibinfo {volume} {5}},\ \bibinfo {pages} {023163} (\bibinfo {year}
  {2023})}\BibitemShut {NoStop}%
\bibitem [{\citenamefont {Peng}\ \emph {et~al.}(2022)\citenamefont {Peng},
  \citenamefont {Salomoni}, \citenamefont {Malinowski}, \citenamefont {Zhang},
  \citenamefont {Hohlfeld}, \citenamefont {Buda-Prejbeanu}, \citenamefont
  {Gorchon}, \citenamefont {Vergès}, \citenamefont {Lin}, \citenamefont
  {Sousa}, \citenamefont {Prejbeanu}, \citenamefont {Mangin},\ and\
  \citenamefont {Hehn}}]{Peng:2022}%
  \BibitemOpen
  \bibfield  {author} {\bibinfo {author} {\bibfnamefont {Y.}~\bibnamefont
  {Peng}}, \bibinfo {author} {\bibfnamefont {D.}~\bibnamefont {Salomoni}},
  \bibinfo {author} {\bibfnamefont {G.}~\bibnamefont {Malinowski}}, \bibinfo
  {author} {\bibfnamefont {W.}~\bibnamefont {Zhang}}, \bibinfo {author}
  {\bibfnamefont {J.}~\bibnamefont {Hohlfeld}}, \bibinfo {author}
  {\bibfnamefont {L.~D.}\ \bibnamefont {Buda-Prejbeanu}}, \bibinfo {author}
  {\bibfnamefont {J.}~\bibnamefont {Gorchon}}, \bibinfo {author} {\bibfnamefont
  {M.}~\bibnamefont {Vergès}}, \bibinfo {author} {\bibfnamefont {J.~X.}\
  \bibnamefont {Lin}}, \bibinfo {author} {\bibfnamefont {R.~C.}\ \bibnamefont
  {Sousa}}, \bibinfo {author} {\bibfnamefont {I.~L.}\ \bibnamefont
  {Prejbeanu}}, \bibinfo {author} {\bibfnamefont {S.}~\bibnamefont {Mangin}}, \
  and\ \bibinfo {author} {\bibfnamefont {M.}~\bibnamefont {Hehn}},\ }\bibfield
  {title} {\enquote {\bibinfo {title} {In plane reorientation induced single
  laser pulse magnetization reversal in rare-earth based multilayer},}\ }\href
  {\doibase 10.48550/ARXIV.2212.13279} {\bibfield  {journal} {\bibinfo
  {journal} {arXiv}\ ,\ \bibinfo {pages} {2212.13279}} (\bibinfo {year}
  {2022})}\BibitemShut {NoStop}%
\bibitem [{\citenamefont {Ceballos}\ \emph {et~al.}(2021)\citenamefont
  {Ceballos}, \citenamefont {Pattabi}, \citenamefont {El-Ghazaly},
  \citenamefont {Ruta}, \citenamefont {Simon}, \citenamefont {Evans},
  \citenamefont {Ostler}, \citenamefont {Chantrell}, \citenamefont {Kennedy},
  \citenamefont {Scott}, \citenamefont {Bokor},\ and\ \citenamefont
  {Hellman}}]{Ceballos:2021}%
  \BibitemOpen
  \bibfield  {author} {\bibinfo {author} {\bibfnamefont {A.}~\bibnamefont
  {Ceballos}}, \bibinfo {author} {\bibfnamefont {A.}~\bibnamefont {Pattabi}},
  \bibinfo {author} {\bibfnamefont {A.}~\bibnamefont {El-Ghazaly}}, \bibinfo
  {author} {\bibfnamefont {S.}~\bibnamefont {Ruta}}, \bibinfo {author}
  {\bibfnamefont {C.~P.}\ \bibnamefont {Simon}}, \bibinfo {author}
  {\bibfnamefont {R.~F.~L.}\ \bibnamefont {Evans}}, \bibinfo {author}
  {\bibfnamefont {T.}~\bibnamefont {Ostler}}, \bibinfo {author} {\bibfnamefont
  {R.~W.}\ \bibnamefont {Chantrell}}, \bibinfo {author} {\bibfnamefont
  {E.}~\bibnamefont {Kennedy}}, \bibinfo {author} {\bibfnamefont
  {M.}~\bibnamefont {Scott}}, \bibinfo {author} {\bibfnamefont
  {J.}~\bibnamefont {Bokor}}, \ and\ \bibinfo {author} {\bibfnamefont
  {F.}~\bibnamefont {Hellman}},\ }\bibfield  {title} {\enquote {\bibinfo
  {title} {Role of element-specific damping in ultrafast, helicity-independent,
  all-optical switching dynamics in amorphous ({Gd},{Tb}){Co} thin films},}\
  }\href {\doibase 10.1103/PhysRevB.103.024438} {\bibfield  {journal} {\bibinfo
   {journal} {Phys. Rev. B}\ }\textbf {\bibinfo {volume} {103}},\ \bibinfo
  {pages} {024438} (\bibinfo {year} {2021})}\BibitemShut {NoStop}%
\bibitem [{\citenamefont {Zhang}\ \emph {et~al.}(2022)\citenamefont {Zhang},
  \citenamefont {Lin}, \citenamefont {Huang}, \citenamefont {Malinowski},
  \citenamefont {Hehn}, \citenamefont {Xu}, \citenamefont {Mangin},\ and\
  \citenamefont {Zhao}}]{Zhang:2022}%
  \BibitemOpen
  \bibfield  {author} {\bibinfo {author} {\bibfnamefont {W.}~\bibnamefont
  {Zhang}}, \bibinfo {author} {\bibfnamefont {J.~X.}\ \bibnamefont {Lin}},
  \bibinfo {author} {\bibfnamefont {T.~X.}\ \bibnamefont {Huang}}, \bibinfo
  {author} {\bibfnamefont {G.}~\bibnamefont {Malinowski}}, \bibinfo {author}
  {\bibfnamefont {M.}~\bibnamefont {Hehn}}, \bibinfo {author} {\bibfnamefont
  {Y.}~\bibnamefont {Xu}}, \bibinfo {author} {\bibfnamefont {S.}~\bibnamefont
  {Mangin}}, \ and\ \bibinfo {author} {\bibfnamefont {W.}~\bibnamefont
  {Zhao}},\ }\bibfield  {title} {\enquote {\bibinfo {title} {Role of
  spin-lattice coupling in ultrafast demagnetization and all optical
  helicity-independent single-shot switching in
  {Gd}$_{1\ensuremath{-}x\ensuremath{-}y}${Tb}$_{y}${Co}$_{x}$ alloys},}\
  }\href {\doibase 10.1103/PhysRevB.105.054410} {\bibfield  {journal} {\bibinfo
   {journal} {Phys. Rev. B}\ }\textbf {\bibinfo {volume} {105}},\ \bibinfo
  {pages} {054410} (\bibinfo {year} {2022})}\BibitemShut {NoStop}%
\bibitem [{\citenamefont {Coey}(1978)}]{Coey:1978}%
  \BibitemOpen
  \bibfield  {author} {\bibinfo {author} {\bibfnamefont {J.~M.~D.}\
  \bibnamefont {Coey}},\ }\bibfield  {title} {\enquote {\bibinfo {title}
  {Amorphous magnetic order},}\ }\href {\doibase 10.1063/1.324880} {\bibfield
  {journal} {\bibinfo  {journal} {J. Appl. Phys.}\ }\textbf {\bibinfo {volume}
  {49}},\ \bibinfo {pages} {1646} (\bibinfo {year} {1978})}\BibitemShut
  {NoStop}%
\bibitem [{\citenamefont {Togami}, \citenamefont {Saito},\ and\ \citenamefont
  {Okamoto}(1986)}]{Togami:1986}%
  \BibitemOpen
  \bibfield  {author} {\bibinfo {author} {\bibfnamefont {Y.}~\bibnamefont
  {Togami}}, \bibinfo {author} {\bibfnamefont {N.}~\bibnamefont {Saito}}, \
  and\ \bibinfo {author} {\bibfnamefont {K.}~\bibnamefont {Okamoto}},\
  }\bibfield  {title} {\enquote {\bibinfo {title} {Anisotropy dispersion and
  its influence on magneto‐optical effect in rare‐earth transition‐metal
  amorphous films},}\ }\href {\doibase 10.1063/1.337577} {\bibfield  {journal}
  {\bibinfo  {journal} {J. Appl. Phys.}\ }\textbf {\bibinfo {volume} {60}},\
  \bibinfo {pages} {3691} (\bibinfo {year} {1986})}\BibitemShut {NoStop}%
\bibitem [{\citenamefont {Wang}\ \emph {et~al.}(2020)\citenamefont {Wang},
  \citenamefont {van Hees}, \citenamefont {Lavrijsen}, \citenamefont {Zhao},\
  and\ \citenamefont {Koopmans}}]{Wang:2020abPL}%
  \BibitemOpen
  \bibfield  {author} {\bibinfo {author} {\bibfnamefont {L.}~\bibnamefont
  {Wang}}, \bibinfo {author} {\bibfnamefont {Y.~L.~W.}\ \bibnamefont {van
  Hees}}, \bibinfo {author} {\bibfnamefont {R.}~\bibnamefont {Lavrijsen}},
  \bibinfo {author} {\bibfnamefont {W.}~\bibnamefont {Zhao}}, \ and\ \bibinfo
  {author} {\bibfnamefont {B.}~\bibnamefont {Koopmans}},\ }\bibfield  {title}
  {\enquote {\bibinfo {title} {Enhanced all-optical switching and domain wall
  velocity in annealed synthetic-ferrimagnetic multilayers},}\ }\href@noop {}
  {\bibfield  {journal} {\bibinfo  {journal} {Appl. Phys. Lett.}\ }\textbf
  {\bibinfo {volume} {117}},\ \bibinfo {pages} {022408} (\bibinfo {year}
  {2020})}\BibitemShut {NoStop}%
\bibitem [{\citenamefont {Li}\ \emph {et~al.}(2022)\citenamefont {Li},
  \citenamefont {van~der Jagt}, \citenamefont {Beens}, \citenamefont
  {Hintermayr}, \citenamefont {Verheijen}, \citenamefont {Bruikman},
  \citenamefont {Barcones}, \citenamefont {Juge}, \citenamefont {Lavrijsen},
  \citenamefont {Ravelosona},\ and\ \citenamefont {Koopmans}}]{Li:2022adPL}%
  \BibitemOpen
  \bibfield  {author} {\bibinfo {author} {\bibfnamefont {P.}~\bibnamefont
  {Li}}, \bibinfo {author} {\bibfnamefont {J.~W.}\ \bibnamefont {van~der
  Jagt}}, \bibinfo {author} {\bibfnamefont {M.}~\bibnamefont {Beens}}, \bibinfo
  {author} {\bibfnamefont {J.}~\bibnamefont {Hintermayr}}, \bibinfo {author}
  {\bibfnamefont {M.~A.}\ \bibnamefont {Verheijen}}, \bibinfo {author}
  {\bibfnamefont {R.}~\bibnamefont {Bruikman}}, \bibinfo {author}
  {\bibfnamefont {B.}~\bibnamefont {Barcones}}, \bibinfo {author}
  {\bibfnamefont {R.}~\bibnamefont {Juge}}, \bibinfo {author} {\bibfnamefont
  {R.}~\bibnamefont {Lavrijsen}}, \bibinfo {author} {\bibfnamefont
  {D.}~\bibnamefont {Ravelosona}}, \ and\ \bibinfo {author} {\bibfnamefont
  {B.}~\bibnamefont {Koopmans}},\ }\bibfield  {title} {\enquote {\bibinfo
  {title} {Enhancing all-optical switching of magnetization by he ion
  irradiation},}\ }\href@noop {} {\bibfield  {journal} {\bibinfo  {journal}
  {Appl. Phys. Lett.}\ }\textbf {\bibinfo {volume} {121}},\ \bibinfo {pages}
  {172404} (\bibinfo {year} {2022})}\BibitemShut {NoStop}%
\bibitem [{\citenamefont {Gorchon}\ \emph {et~al.}(2016)\citenamefont
  {Gorchon}, \citenamefont {Wilson}, \citenamefont {Yang}, \citenamefont
  {Pattabi}, \citenamefont {Chen}, \citenamefont {He}, \citenamefont {Wang},
  \citenamefont {Li},\ and\ \citenamefont {Bokor}}]{Gorchon:2016}%
  \BibitemOpen
  \bibfield  {author} {\bibinfo {author} {\bibfnamefont {J.}~\bibnamefont
  {Gorchon}}, \bibinfo {author} {\bibfnamefont {R.~B.}\ \bibnamefont {Wilson}},
  \bibinfo {author} {\bibfnamefont {Y.}~\bibnamefont {Yang}}, \bibinfo {author}
  {\bibfnamefont {A.}~\bibnamefont {Pattabi}}, \bibinfo {author} {\bibfnamefont
  {J.~Y.}\ \bibnamefont {Chen}}, \bibinfo {author} {\bibfnamefont
  {L.}~\bibnamefont {He}}, \bibinfo {author} {\bibfnamefont {J.~P.}\
  \bibnamefont {Wang}}, \bibinfo {author} {\bibfnamefont {M.}~\bibnamefont
  {Li}}, \ and\ \bibinfo {author} {\bibfnamefont {J.}~\bibnamefont {Bokor}},\
  }\bibfield  {title} {\enquote {\bibinfo {title} {Role of electron and phonon
  temperatures in the helicity-independent all-optical switching of
  {GdFeCo}},}\ }\href {\doibase 10.1103/PhysRevB.94.184406} {\bibfield
  {journal} {\bibinfo  {journal} {Phys. Rev. B}\ }\textbf {\bibinfo {volume}
  {94}},\ \bibinfo {pages} {184406} (\bibinfo {year} {2016})}\BibitemShut
  {NoStop}%
\bibitem [{\citenamefont {Lalieu}\ \emph {et~al.}(2017)\citenamefont {Lalieu},
  \citenamefont {Peeters}, \citenamefont {Haenen}, \citenamefont {Lavrijsen},\
  and\ \citenamefont {Koopmans}}]{Lalieu:2017}%
  \BibitemOpen
  \bibfield  {author} {\bibinfo {author} {\bibfnamefont {M.~L.~M.}\
  \bibnamefont {Lalieu}}, \bibinfo {author} {\bibfnamefont {M.~J.~G.}\
  \bibnamefont {Peeters}}, \bibinfo {author} {\bibfnamefont {S.~R.~R.}\
  \bibnamefont {Haenen}}, \bibinfo {author} {\bibfnamefont {R.}~\bibnamefont
  {Lavrijsen}}, \ and\ \bibinfo {author} {\bibfnamefont {B.}~\bibnamefont
  {Koopmans}},\ }\bibfield  {title} {\enquote {\bibinfo {title} {Deterministic
  all-optical switching of synthetic ferrimagnets using single femtosecond
  laser pulses},}\ }\href {\doibase 10.1103/PhysRevB.96.220411} {\bibfield
  {journal} {\bibinfo  {journal} {Phys. Rev. B}\ }\textbf {\bibinfo {volume}
  {96}},\ \bibinfo {pages} {220411} (\bibinfo {year} {2017})}\BibitemShut
  {NoStop}%
\bibitem [{\citenamefont {Peeters}, \citenamefont {van Ballegooie},\ and\
  \citenamefont {Koopmans}(2022)}]{Peeters:2022}%
  \BibitemOpen
  \bibfield  {author} {\bibinfo {author} {\bibfnamefont {M.~J.~G.}\
  \bibnamefont {Peeters}}, \bibinfo {author} {\bibfnamefont {Y.~M.}\
  \bibnamefont {van Ballegooie}}, \ and\ \bibinfo {author} {\bibfnamefont
  {B.}~\bibnamefont {Koopmans}},\ }\bibfield  {title} {\enquote {\bibinfo
  {title} {Influence of magnetic fields on ultrafast laser-induced switching
  dynamics in {Co/Gd} bilayers},}\ }\href {\doibase
  10.1103/PhysRevB.105.014429} {\bibfield  {journal} {\bibinfo  {journal}
  {Phys. Rev. B}\ }\textbf {\bibinfo {volume} {105}},\ \bibinfo {pages}
  {014429} (\bibinfo {year} {2022})}\BibitemShut {NoStop}%
\end{thebibliography}%

\end{document}